\documentclass[letterpaper]{article}

\usepackage{times}
\usepackage{helvet}
\usepackage{courier}
\usepackage[utf8]{inputenc}
\usepackage[english]{babel}
\usepackage{amsmath}
\usepackage{amsthm}
\usepackage{amssymb}
\usepackage{url}
\usepackage{graphicx}
\usepackage{array}
\usepackage{float}

\begin{document}

\title{A data-driven analysis to question epidemic models for citation cascades on the blogosphere}

\author{
Abdelhamid Salah Brahim\\
LIAFA, Universit\'e Paris 7 Denis Diderot\\
salah@liafa.jussieu.fr
\and
Lionel Tabourier\\
naXys, Universit\'e de Namur\\
lionel.tabourier@fundp.ac.be
\and
Bénédicte Le Grand\\
CRI, Universit\'e Paris 1 Panth\'eon-Sorbonne\\
benedicte.le-grand@univ-paris1.fr
}

\maketitle

\begin{abstract}

Citation cascades in blog networks are often considered as traces of information spreading on this social medium. 
In this work, we question this point of view using both a structural and semantic analysis of five months activity of the most representative blogs of the french-speaking community.
Statistical measures reveal that our dataset shares many features with those that can be found in the literature, suggesting the existence of an identical underlying process.
However, a closer analysis of the post content indicates that the popular epidemic-like descriptions of cascades are misleading in this context.
A basic model, taking only into account the behavior of bloggers and their restricted social network, accounts for several important statistical features of the data.
These arguments support the idea that citations primary goal may not be information spreading on the blogosphere.\\

\noindent
\textbf{Keywords:}
blog network,
citation cascades,
information spreading,
statistical analysis,
epidemic models
%

\end{abstract}


\section{Introduction}

During the last decade, the \textit{World Wide Web} functions and usages have been widely impacted by the popularization of user-friendly content editors, most noticeably wikis and blogging platforms.
Blogs in particular emerged as a public space for opinion broadcasting as well as a form of participative journalism or even a shop-window for commercial activities.

This new kind of media has focused much attention from the scientific community.
In addition to the novelty of the phenomenon, the blogosphere is a huge dataset of rich and publicly available content, and thus comes across as a means to understand the principles of information spreading on social networks.
The seemingly similar mechanisms at stake in epidemiology and information adoption made the tools of the former a popular source of inspiration.
Both measurements and models have been developed in this line, often describing the information transmission as an infection-like process.

In this article we question the legitimacy of these approaches as a proper trace of information spreading, in the specific context of bloggers citing each other.
Following a typical complex networks approach, we use statistical tools to describe large dynamical datasets, and observe some similarities with an unrelated dataset of the literature.
We put forward the idea that usual measures seize mostly effects which stem from individual posting and citing behavior of bloggers.
This assumption is given more credit by a content-oriented analysis, which helps to understand the origin of the properties captured by these statistical tools.
Finally, we provide a content-independent model to further support this hypothesis; it also yields clues on where an observer should look for traces of information spreading in this dataset.

\subsection{Related work}

The rise of blogs among online social media has been discussed from a sociological point of view in numerous papers, e.g., \cite{lenhart2006bloggers} give some insights about bloggers demographics and cultural behaviors; in \cite{gill2004can}, the author describes their influence on society. 
Given the size and richness of the blog datasets, automatic classification and text-mining tools have been widely used to study the dynamics of trends and opinions in the blogosphere \cite{gruhl2004information,adar2005tracking,kumar2005bursty,joshi2007web,macskassy2011contextual}.
For example, some studies concentrate on the political blogosphere to understand the ties between political parties, in particular the way information spreads from a group to another \cite{adamic2005political,cointet2009socio}.
The question of trends is closely related to the definition of authority, influence and trust in social networks~\cite{gill2004can,cointet2009socio}.
As such, these text-mining tools are also used for various practical purposes as online advertising~\cite{stewart2007discovering} or search engines, ranking blogs according to their spreading ability~\cite{adar2004implicit}.

A large amount of works has been dedicated to discovering routes of information spreading in online social media such as products recommendation platforms \cite{leskovec2006patterns}, online games \cite{bakshy2009social}, Flickr \cite{cha2009measurement}, Twitter \cite{romero2011differences,myers2012information}, and of course the blogosphere. 
These routes are often inferred from the structure of the social network and the dynamical behavior of the agents (adopting a new technology, joining a group), especially when there is no concrete trace of the peer influence, e.g., \cite{adar2005tracking,cha2009measurement}.
However, the observed contagion may be the result of direct adoption from one's neighbor, but also a consequence of homophily, as connected people are prone to behave similarly~\cite{aral2009distinguishing,bakshy2012role}, making route inference methods questionable.

In this work, we use data in which the connection between users is explicit and combine it with a popular cascade-like description of the spreading~\cite{adar2005tracking,leskovec2007cascading,mcglohon2007finding,bakshy2009social,gotz2009modeling,li2009blog,papagelis2009information}.
More precisely, we adopt definitions very similar to the ones developed in \cite{leskovec2007cascading}, and will therefore often refer to this study for comparison purposes.
In addition to the statistical analyses of these datasets, models have been proposed to explain the observed features \cite{gruhl2004information,kumar2005bursty,leskovec2007cascading,gotz2009modeling}.
Among them, many are inspired by virus spreading models in epidemiology; in the following, we address the issue of the relevance of this particular class of models in the context of blog networks.

\subsection{Outline}

Section~\ref{sec:data} is dedicated to the exposition of the studied blog dataset.
After describing how the data have been collected, we achieve standard statistical measurements with frequent comparisons to similar measures in~\cite{leskovec2007cascading}.
We then investigate the contents of a subset of posts both qualitatively and quantitatively, in regards to spreading processes phenomena.
In Section~\ref{sec:model}, we propose a simple model of the data to mimic the statistical features aforementioned.
It brings us to challenge usual depictions of information propagation in the blogosphere, and more generally on online social media.

\section{Dataset \label{sec:data}}

\subsection{Description and filtering}

The data corpus analyzed in this paper has been obtained by daily crawls of 10,309 blogs during five months (from February $1^{st}$ to July $1^{st}$ 2010), yielding 848,026 posts.
These blogs have been selected according to their popularity, activity and their representativeness of the French-speaking blogosphere.
It therefore excludes blogs which, even if public, aim at informing a small group of friends (such as teenage blogs of the \url{skyrock.com} platform, popular in France).
These \textit{A-list} blogs have been crawled by Linkfluence, a company specialized in online opinion watch (\url{linkfluence.net}) during a research project called \textit{Webfluence}.
In the following, it will thus be referred to as \textit{Webfluence} dataset.

We focus here on \emph{citation links}: consider a post $P_a$ from blog A and a post $P_b$ from blog B. 
If $P_a$ contains the URL of $P_b$, then there is a citation link from $P_a$ to $P_b$.
Notice that other kinds of interaction (e.g. comments, blogrolls) are not taken into account here.
We collected 1,079,195 citation links in this dataset, but after filtering out citations pointing to posts outside the dataset, citations from a blog to itself\footnote{The goal of this paper is to study diffusion phenomena and such citations do not contribute to the spreading process.} and crawling artifacts (e.g. citation from an anterior post or different indexations of a same post), this came down to 3,199 blogs publishing 461,134 posts\footnote{Among them 24,938 are involved as source or destination in citation links contained in the crawling period.} and 20,885 citation links, for which both source and destination are in the crawling period.

\subsection{Information items and routes}

Various hypotheses have been made to track information circulating on the blogosphere.
Most of them rely on the assumption that diffusion can be described as a piece of information moving on a network connecting blogs.  
For the sake of clarity, we will call this (hypothetical) element of information \emph{item} and a \emph{route} is the path followed by such an item.

In \cite{adar2004implicit,adar2005tracking,cointet2009socio}, the authors chose to consider the reference to a particular resource (picture, URL, ...) as an item.
Such a definition is unambiguous, but there is in general no guarantee about where a member of the cascade has found the information, in other words the route is unknown.
The authors of \cite{adar2004implicit,adar2005tracking} looked for subtle criteria to learn the routes from past behaviors, however their studies suppose that routes remain in the blog network, as opposed to information brought by other media.
In \cite{cointet2009socio}, this point of view is refined by only taking into account resources associated to the citation of the source post, but it reduces drastically the set of usable citations in the dataset.

Here, we explore the approach developed in \cite{gotz2009modeling}.
The authors consider that a citation from a post to another is a possible path for information spreading. 
We investigate the definition of an item propagating through such a route, i.e. we are looking for a piece of information that would remain unchanged when a blog is citing another. 
As we will show in this paper through our observations, in general it is not possible to define such an item.

\subsection{Cascades of citations}

\begin{figure}[h!]
\begin{center}
\includegraphics[width=.5\linewidth]{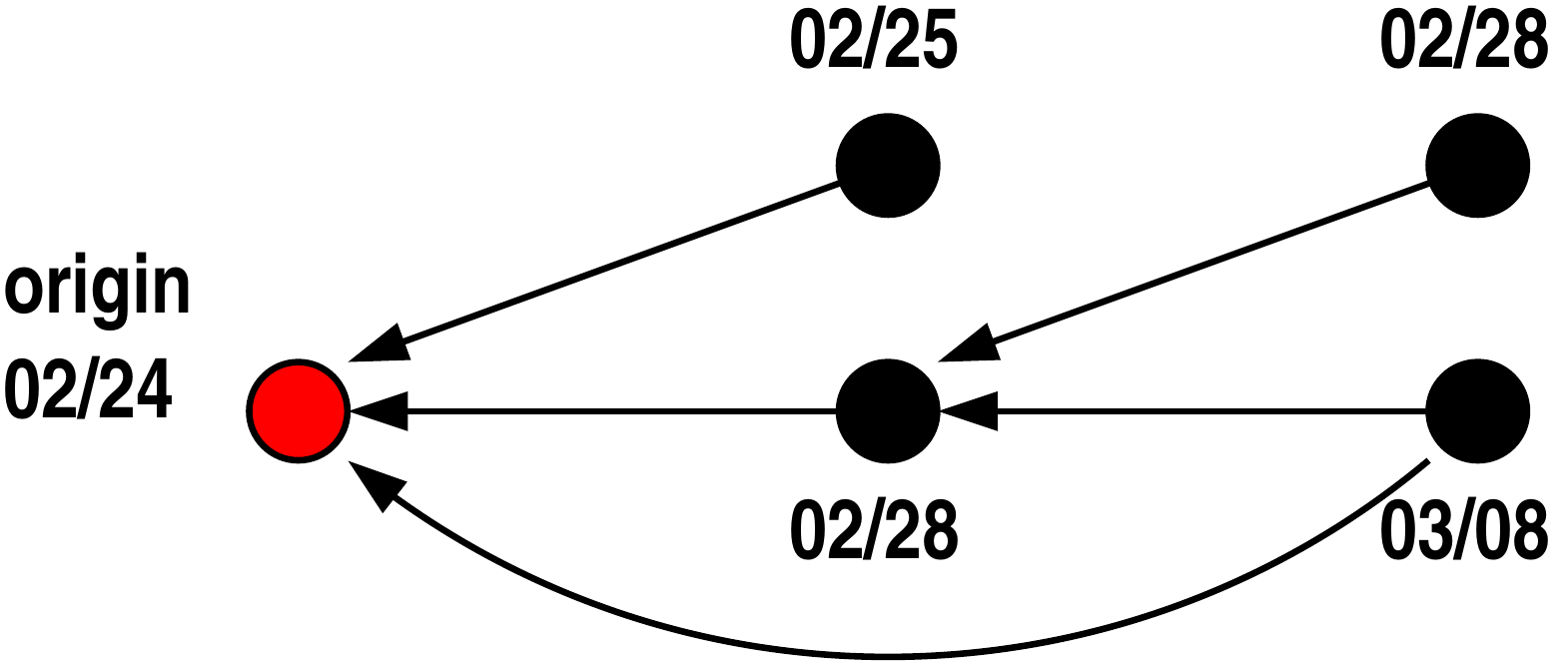}
\end{center}
\caption{\label{fig:ex_cascade}
Example of cascade. 
Here, a node is a post and an arc is a citation link.
The subgraph obtained is a DAG.}
\end{figure}

Describing a route as a \emph{cascade} is now a quite usual picture of information spreading on social networks \cite{adar2005tracking,leskovec2007cascading,gotz2009modeling,li2009blog,papagelis2009information}.
In our study, a cascade is a subgraph of the post network where nodes correspond to posts and edges to citation links. 
It is built in the following way: 
its first node is a post with no outgoing link (the \textit{origin}), 
posts citing it are included in the subgraph with the corresponding directed citation link,
then we look for posts citing these posts in the dataset, and the process carries on recursively.
Because of the temporal ordering of citations, such a definition implies that cascades are directed acyclic graphs (DAG); an example of cascade is given in Fig~\ref{fig:ex_cascade}.
According to this definition, we detect $10,667$ non-trivial cascades\footnote{i.e., not isolated posts} in the \textit{Webfluence} dataset.

In a former paper, the structural properties of these cascades have been discussed relatively to a predefined community hierarchy~\cite{salahbrahim2012diffusion}.
We will now investigate in more details the cascades with regard to their content and to the way information may move from a post to another.
It is worth noticing too that this definition is identical to the one in \cite{leskovec2007cascading} and \cite{mcglohon2007finding}, allowing us to compare the features of other datasets to ours. 
Another possible convention consists in merging all cascades which feature the same post, see for example~\cite{li2009blog}.

\subsection{Comparison to existing results}

We compare standard measurements on the \textit{Webfluence} dataset to the features observed in \cite{leskovec2007cascading}, where the authors analyzed a 90-days crawl of about 45,000 active blogs connected by 205,000 citations in 2005 --- thus larger than ours.
So, considering the period and the fact that the French-speaking community is only a small part of the world-wide blogosphere, these datasets are unrelated. 
Yet, both datasets have been processed in a similar way in order to identify general properties of the blog networks.

\subsubsection{Citation features}

\begin{figure}[h!]
\begin{center}
\includegraphics[angle=-90,width=.49\linewidth]{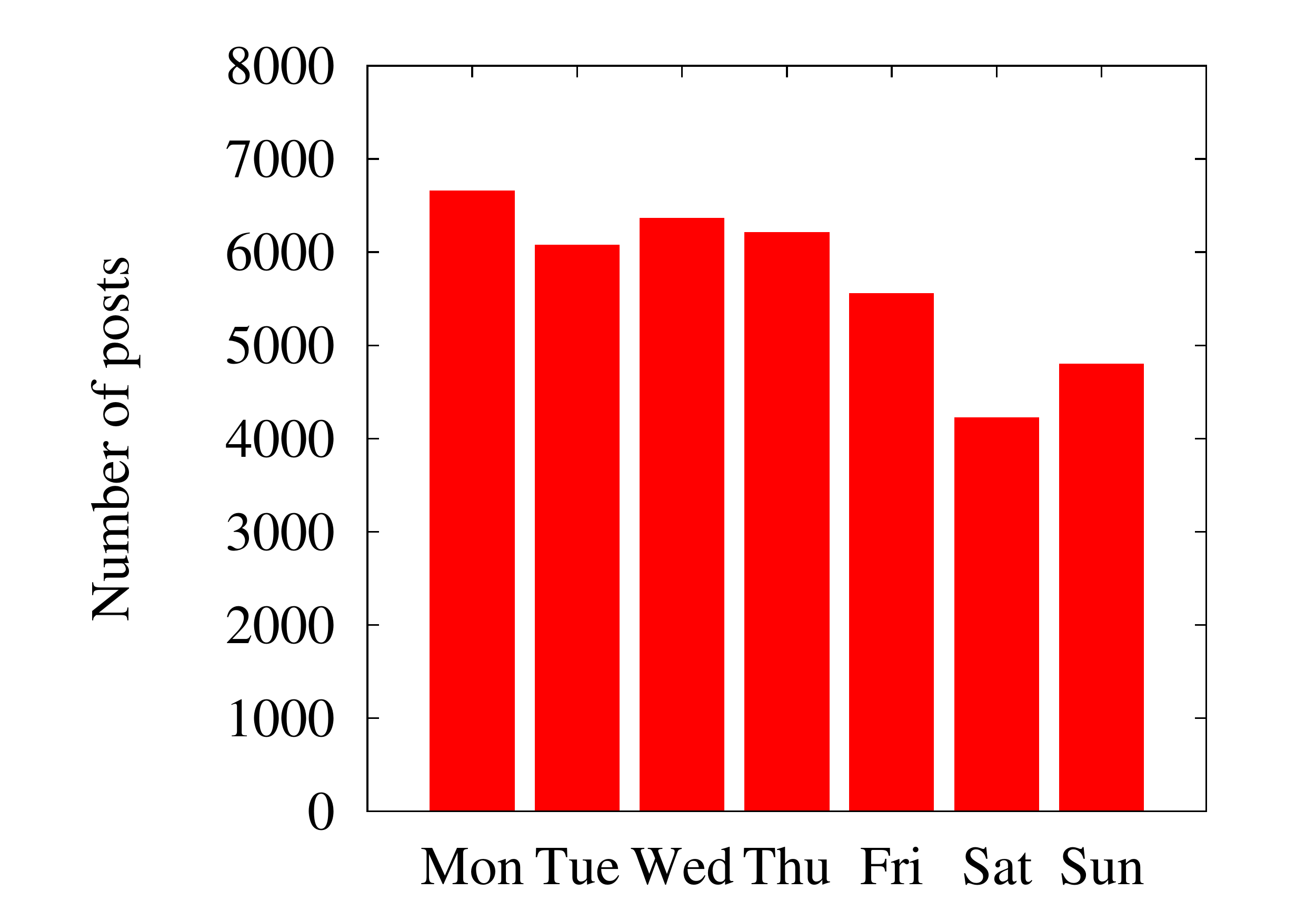}
\includegraphics[angle=-90,width=.49\linewidth]{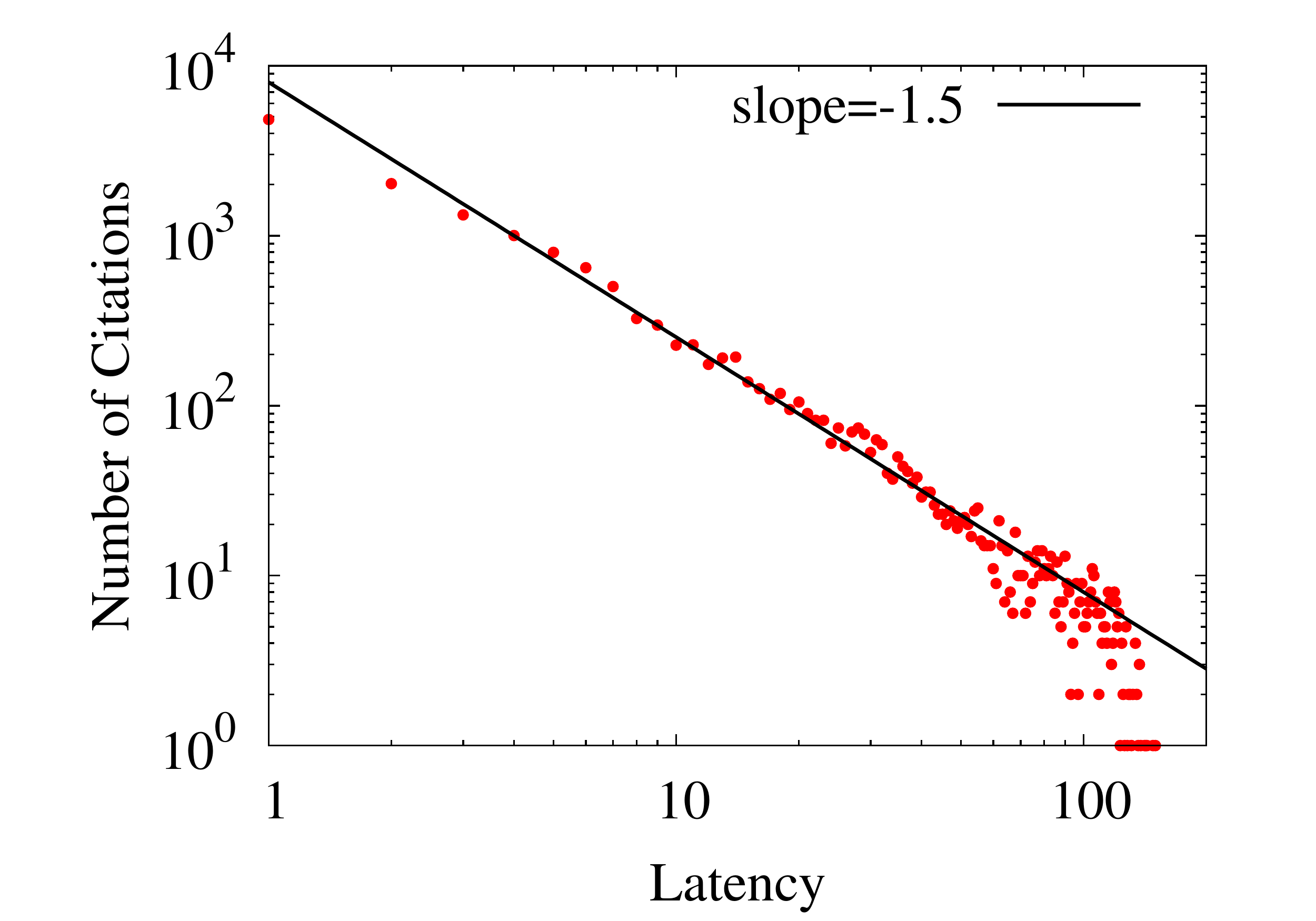}
\end{center}
\caption{\label{fig:cit_dyn}
Activity and citation dynamics statistics.
Left: average number of posts published per day of the week. 
Right: distribution of latencies (temporal gaps) between a post and its citations.
}
\end{figure}

First, we plot on Fig.~\ref{fig:cit_dyn} some features characterizing our dataset's activity and citation dynamics during the crawling period.

The number of posts published as a function of the day of the week exhibits a usual behavior: bloggers publish less during the week-end, reducing their activity by about 27\% when compared to another day of the week (around 40\% in \cite{leskovec2007cascading}).

The distribution of latencies between a post and its citations looks roughly like a power-law with a cut-off.
The cut-off is known to be an effect of the finite time-window, as observed for example in \cite{vazquez2007impact} in the context of email networks; the slope of the exponent is close to -1.5 (-1.6 in \cite{leskovec2007cascading}).
Notice that this shape is modulated by a week-cycle: for example it is slightly more likely than expected by the model to be cited within 7 days, and slightly less to be cited within 8 days.\\

\begin{figure}[h!]
\begin{center}
\includegraphics[angle=-90,width=.49\linewidth]{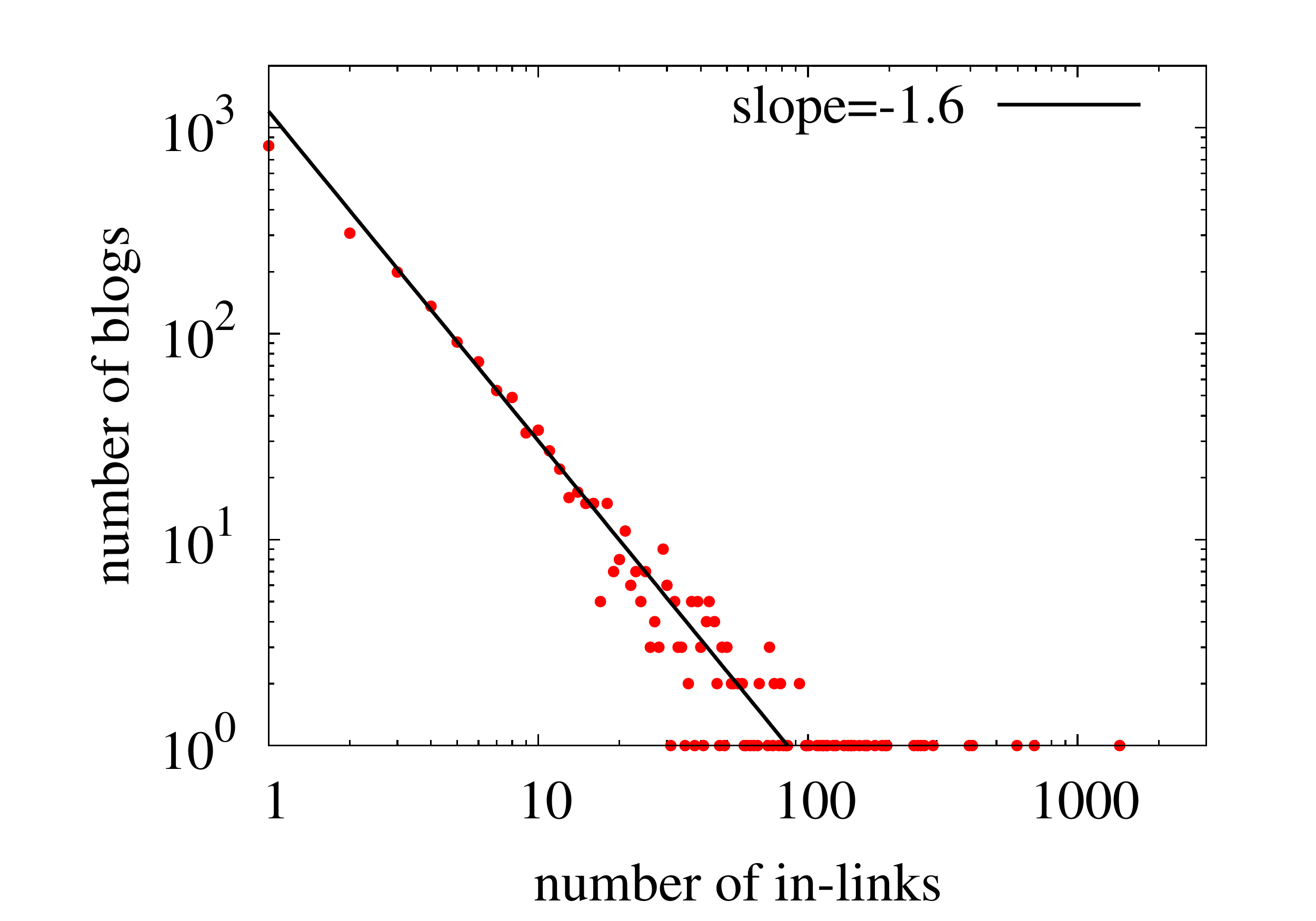}
\includegraphics[angle=-90,width=.49\linewidth]{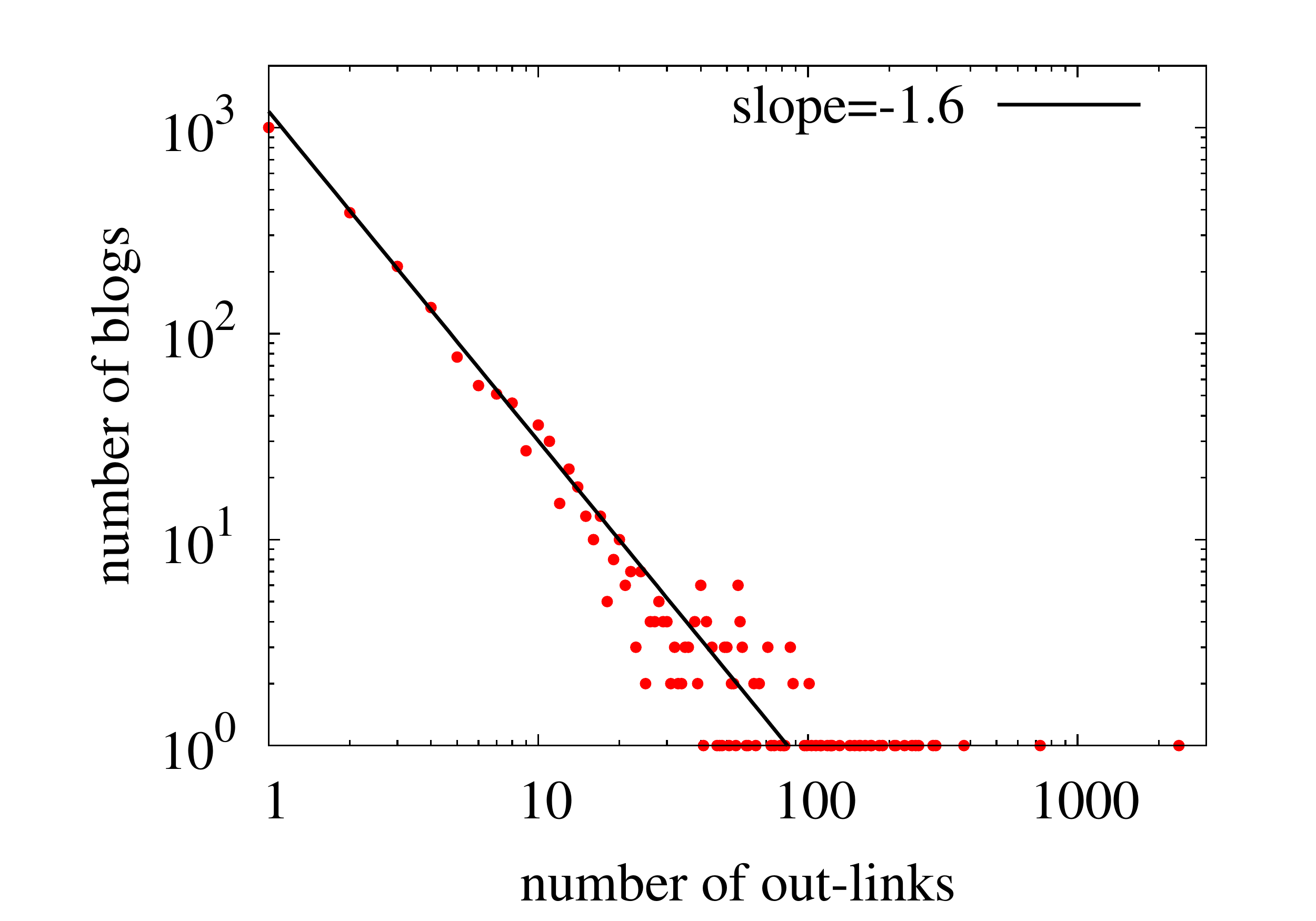}
\includegraphics[angle=-90,width=.49\linewidth]{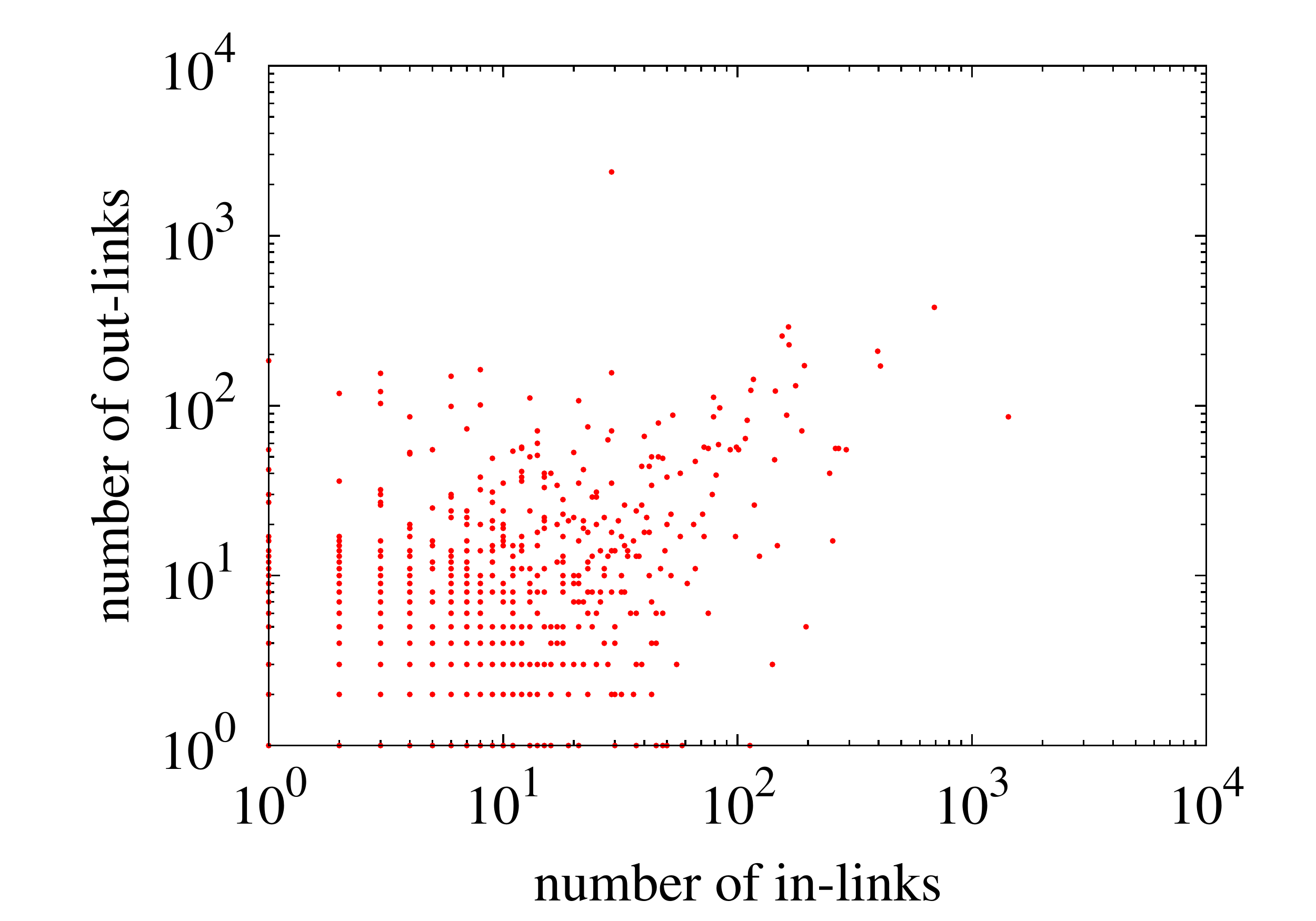}
\end{center}
\caption{\label{fig:deg_dist}
Blog network statistics.
Top left: distribution of incoming links per blog.
Top right: distribution of outgoing links per blog.
Bottom: number of in-links as a function of the number of out-links.
}
\end{figure}

Then, we report on Fig.~\ref{fig:deg_dist} the distributions of in-coming and out-going links for each blog, as well as the correlation between them.
Both qualitative and quantitative behaviors are quite similar to the ones in \cite{leskovec2007cascading}:
degree distributions are heavy-tailed, and may be approximated by power-laws --- even if a 1 to 100 range is not sufficient to be affirmative about the quality of such a model.
The exponents measured are quantitatively close too, as summarized in Table~\ref{tab:comparative_res}; however, the slope is not as steep as other values reported in the literature, which is not surprising, since the steepening is known to vary with the time-window of measurement~\cite{shi2007looking}.
The Pearson correlation coefficient computed from these distributions is about 0.18 (0.16 in \cite{leskovec2007cascading}).
If we consider that the number of citations is an appropriate measure of attention on the web, the value supports the idea that attention and activity are not much correlated in the blogosphere.

\subsubsection{Cascade features}

\begin{figure}[h!]
\begin{center}
\includegraphics[angle=-90,width=.49\linewidth]{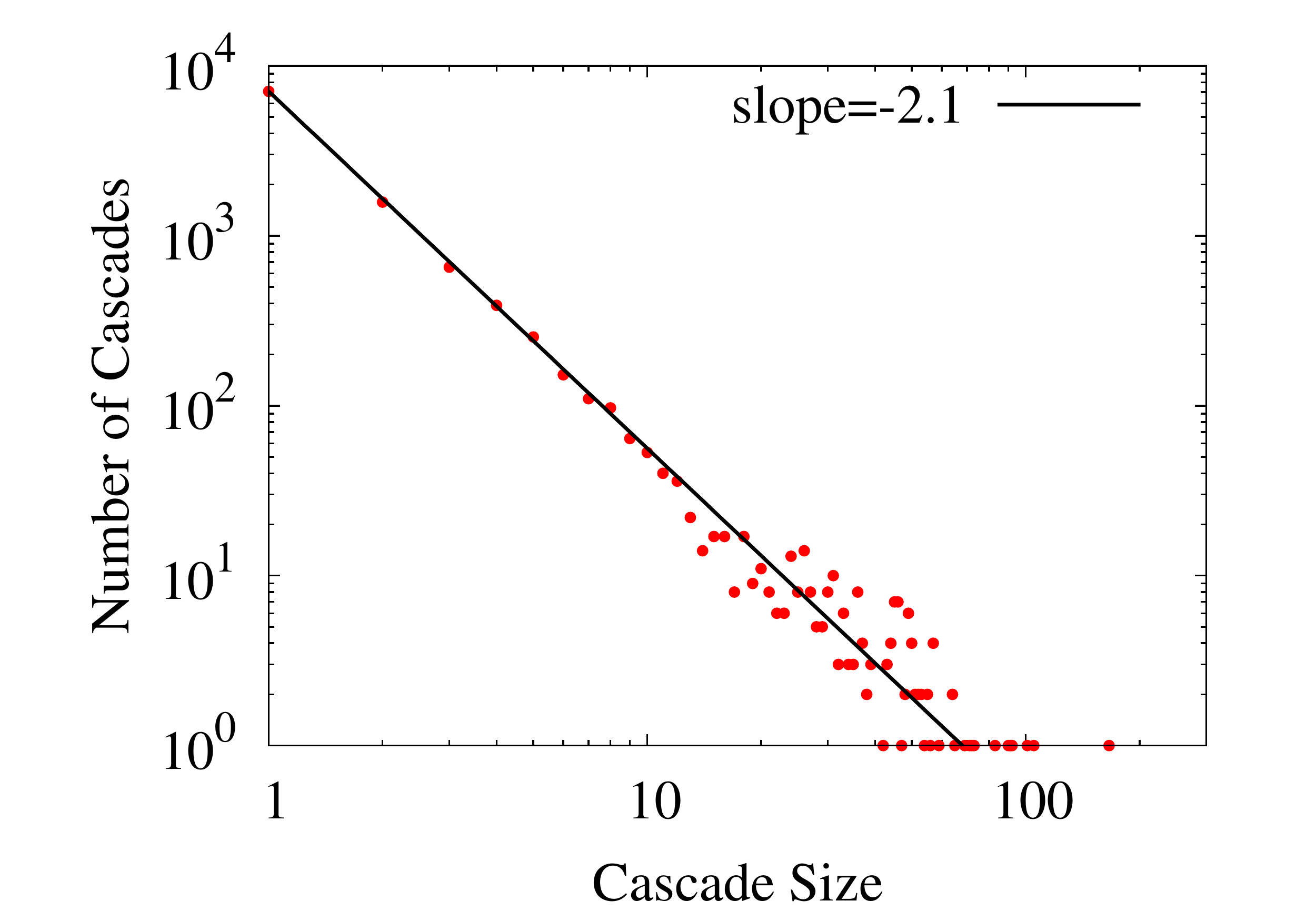}
\includegraphics[angle=-90,width=.49\linewidth]{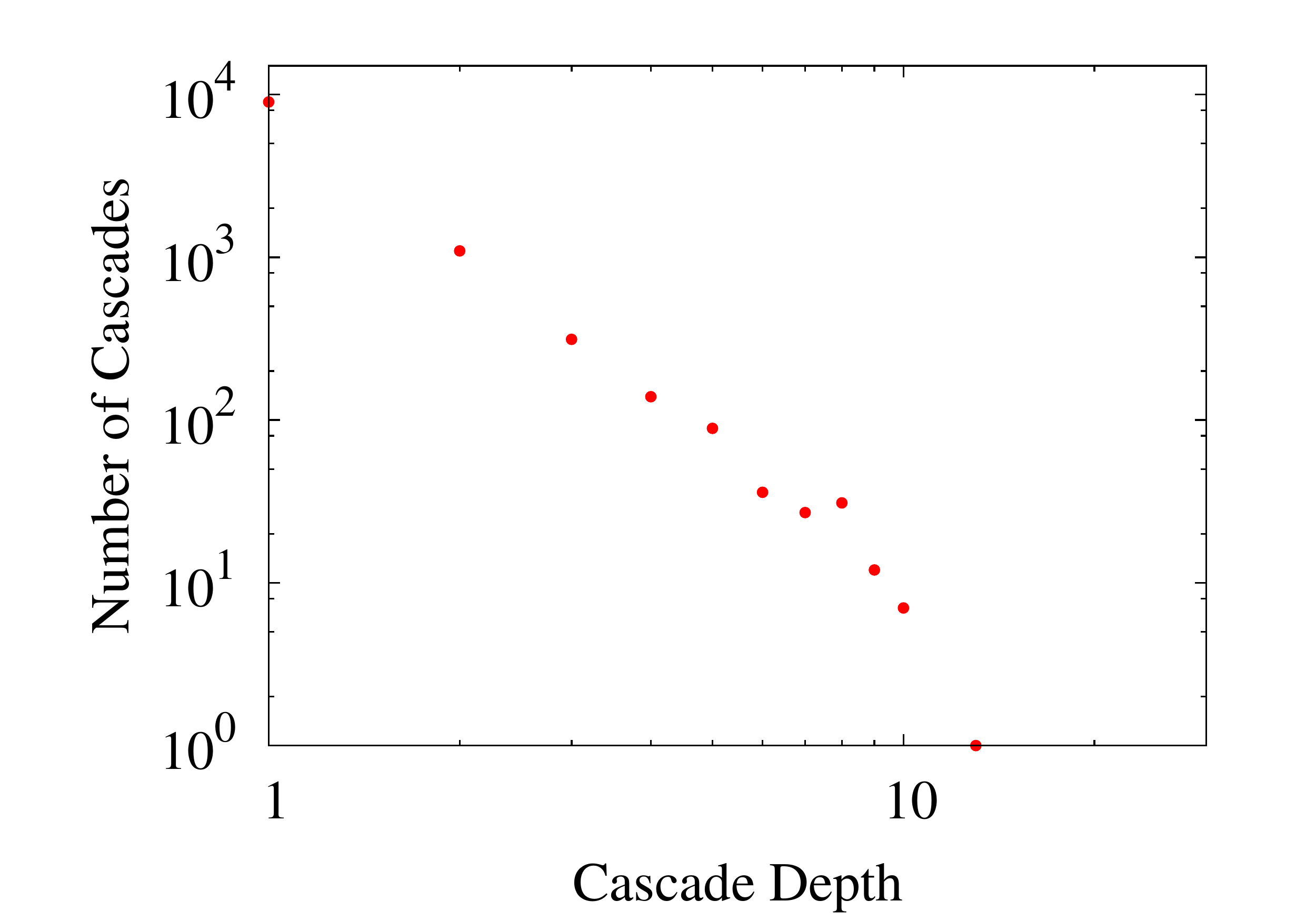}
\end{center}
\caption{\label{fig:casc_size}
Left: distribution of cascade sizes.
Right: distribution of cascade depths.
}
\end{figure}

%
%
We plot the distribution of cascade sizes --- the size is the number of posts in the cascade, origin excluded.
It may be fitted by a power-law model with a slope close to the one in \cite{leskovec2007cascading}.
We considered the cascade depth too, defined here as the longest distance\footnote{The distance between two nodes is defined as the number of arcs of the shortest directed path from a node to the other} between any node of the cascade to its origin (Figure~\ref{fig:casc_size}).
%

We then investigate cascades shapes by ranking them in a decreasing order according to their frequency in the dataset.
Strikingly enough, we can see in Table~\ref{tab:shape_casc} that the ranking of the most common shapes is very close to the hierarchy detailed in \cite{leskovec2008thesis} --- it should be interpreted cautiously with regard to the low number of patterns measured.
Note that at equal size, star-shaped cascades (like the ones ranked 2, 3 or 5) are more frequent than chain-shaped (4 and 12).
It is usual to compare the amount of patterns to a suitable random model to evaluate which ones are over- and under-represented; this point will be developed in Section~\ref{sec:model}.



\begin{table*}[ht]
\begin{center}
\begin{tabular}{|>{\centering}m{2cm}||>{\centering}m{2cm}|>{\centering}m{2cm}|>{\centering}m{2cm}|>{\centering}m{2cm}|>{\centering\arraybackslash}m{2cm}|}
\hline
Rank & 1 & 2 & 3 & 4 & 5 \\
\hline
Rank in \cite{leskovec2008thesis} & 1 & 2 & 3 & 4 & 5 \\
\hline
Shape & 
\includegraphics[angle=90,width=1cm]{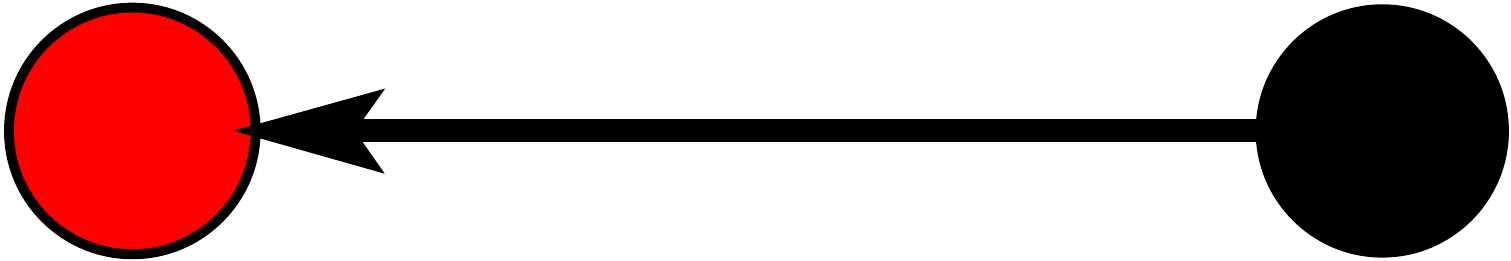} & 
\includegraphics[angle=90,width=1cm]{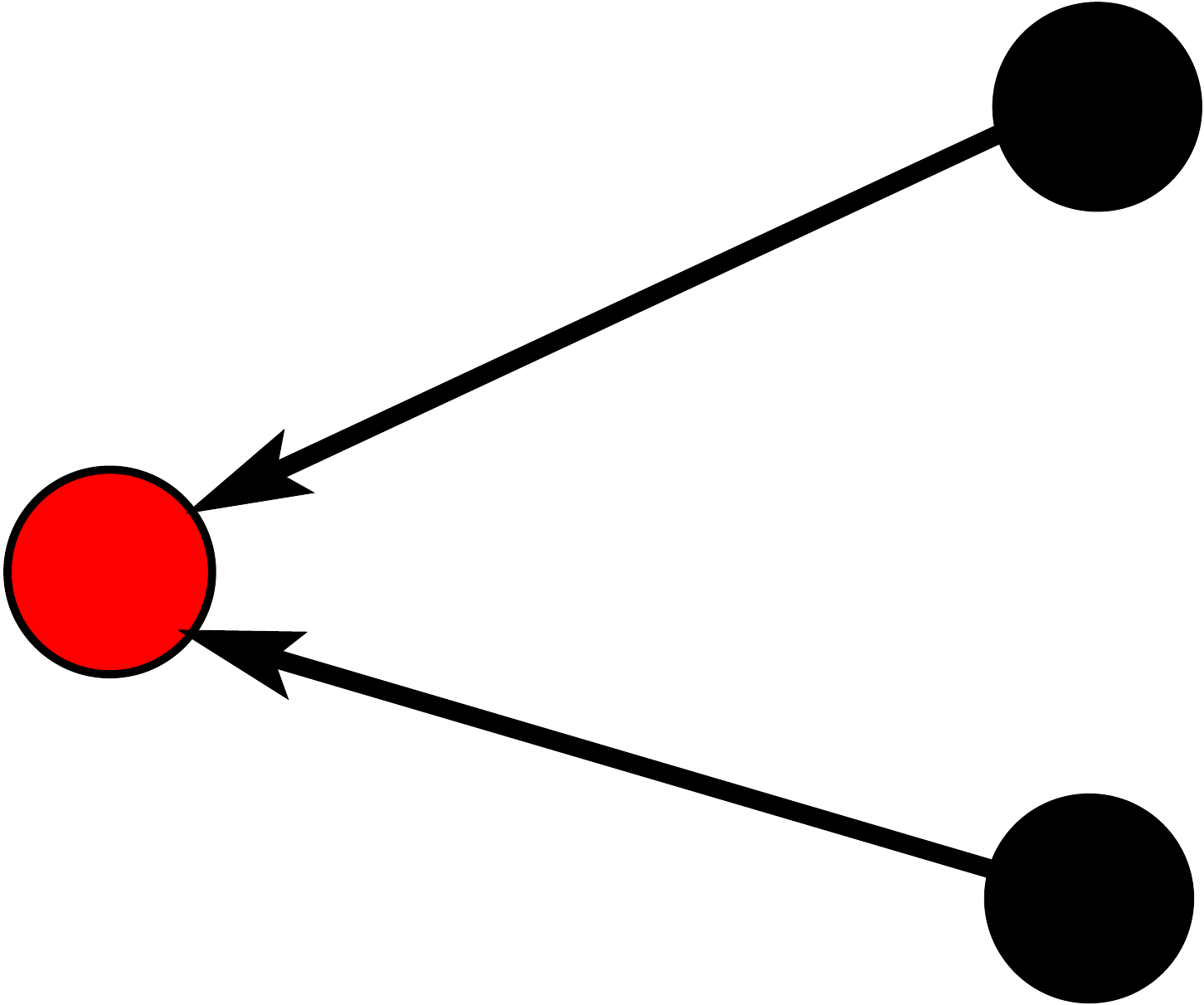} &
\includegraphics[angle=90,width=1cm]{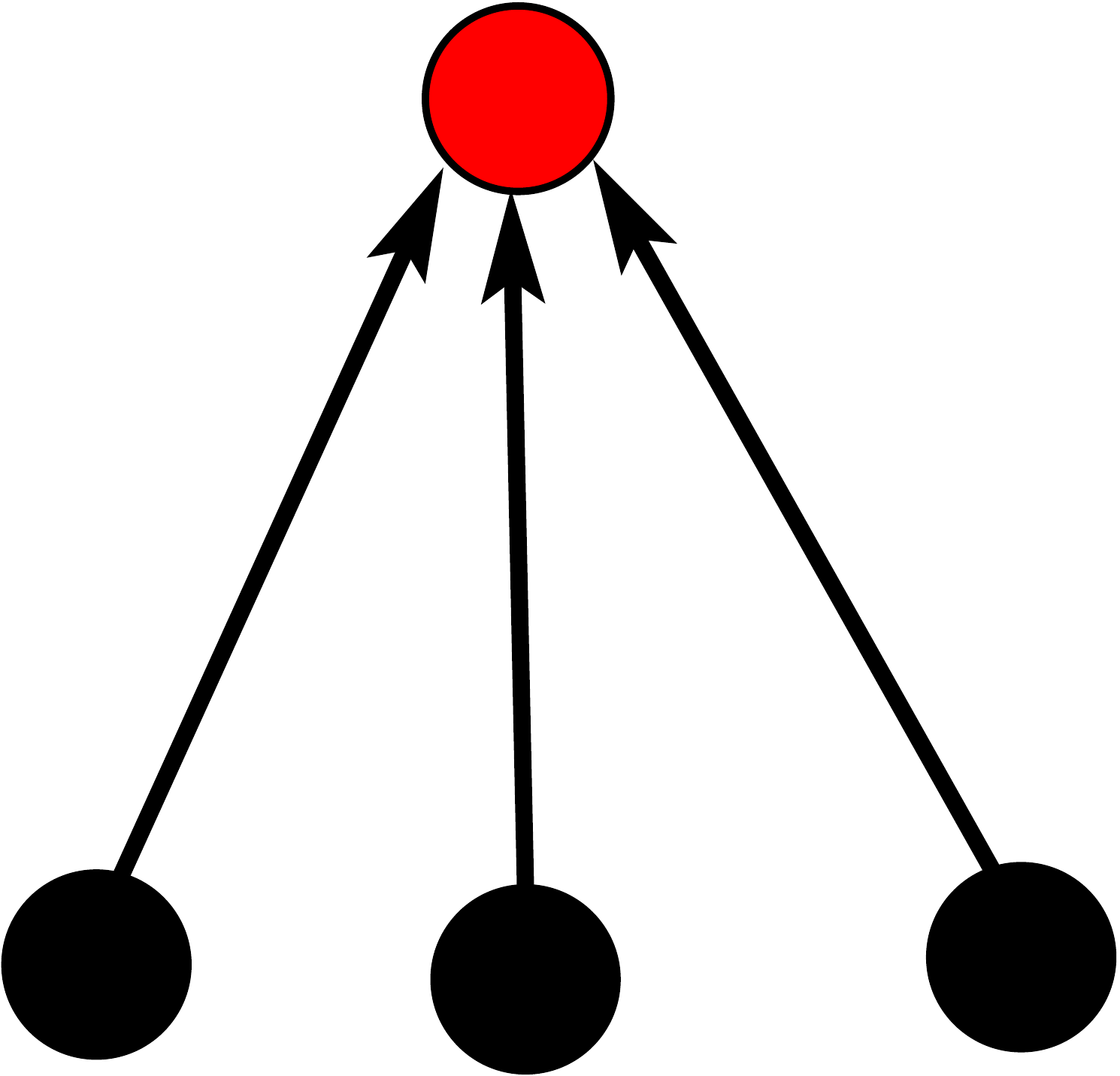} &
\includegraphics[width=2cm]{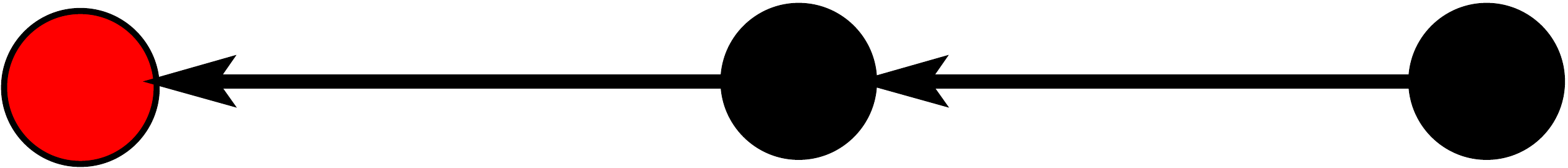} &
\includegraphics[width=1cm]{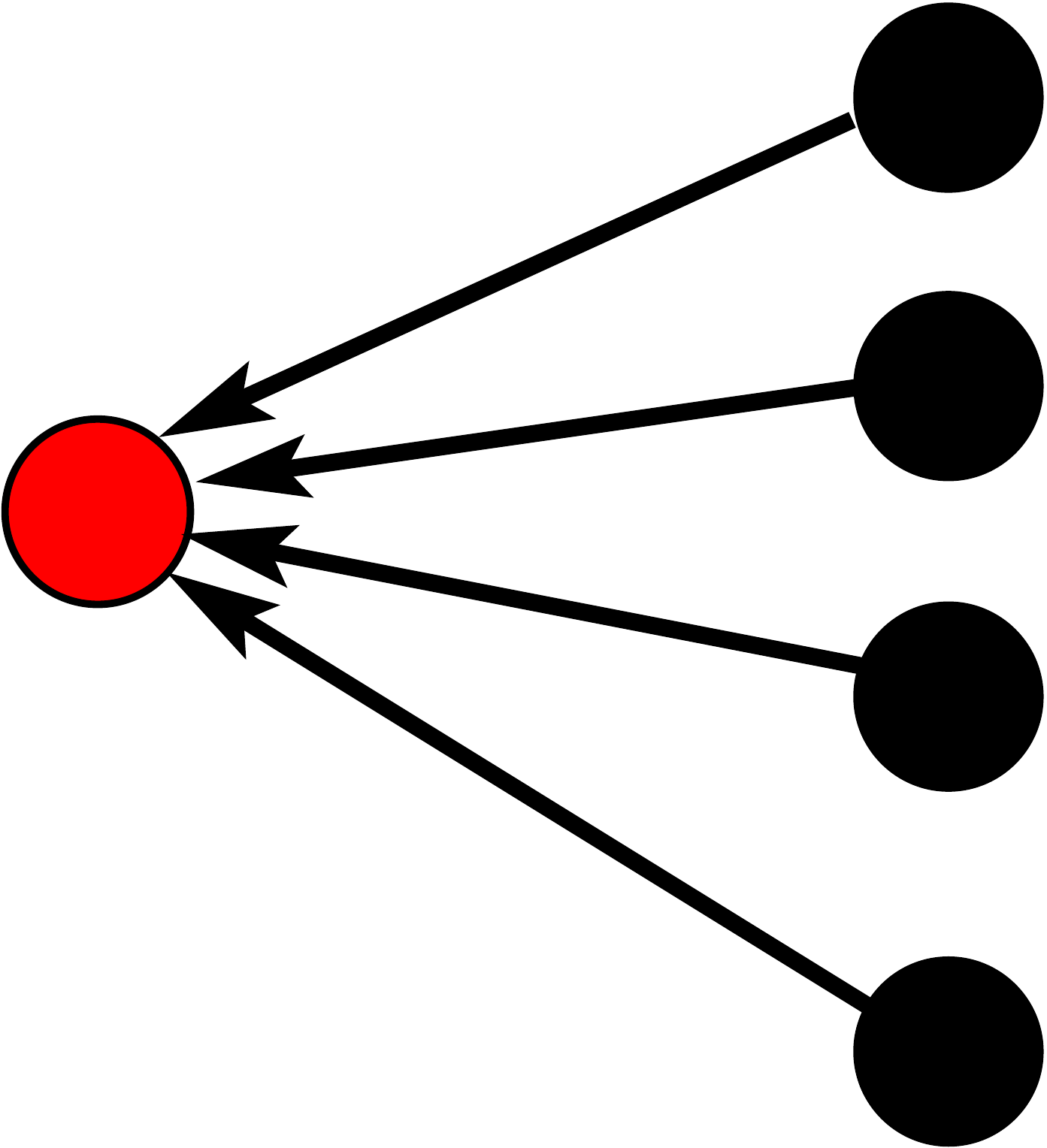} \\
\hline
Frequency & 6992 & 1173 & 397 & 370 & 182 \\
\hline
\end{tabular}

\vspace{2mm}

\begin{tabular}{|>{\centering}m{2cm}||>{\centering}m{2cm}|>{\centering}m{2cm}|>{\centering}m{2cm}|>{\centering}m{2cm}|>{\centering\arraybackslash}m{2cm}|}
\hline
Rank & 6 & 7 & 8 & 9 & 10 \\
\hline
Rank in \cite{leskovec2008thesis} & 6 & 7 & 10 & 11 & 8 \\
\hline
Shape & 
\includegraphics[width=2cm]{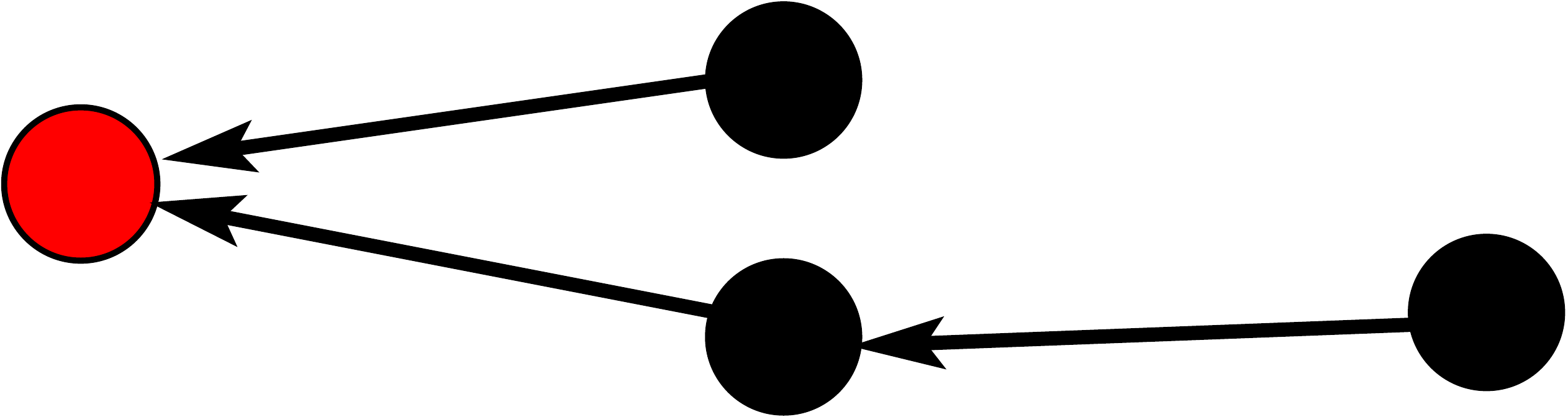} & 
\includegraphics[width=2cm]{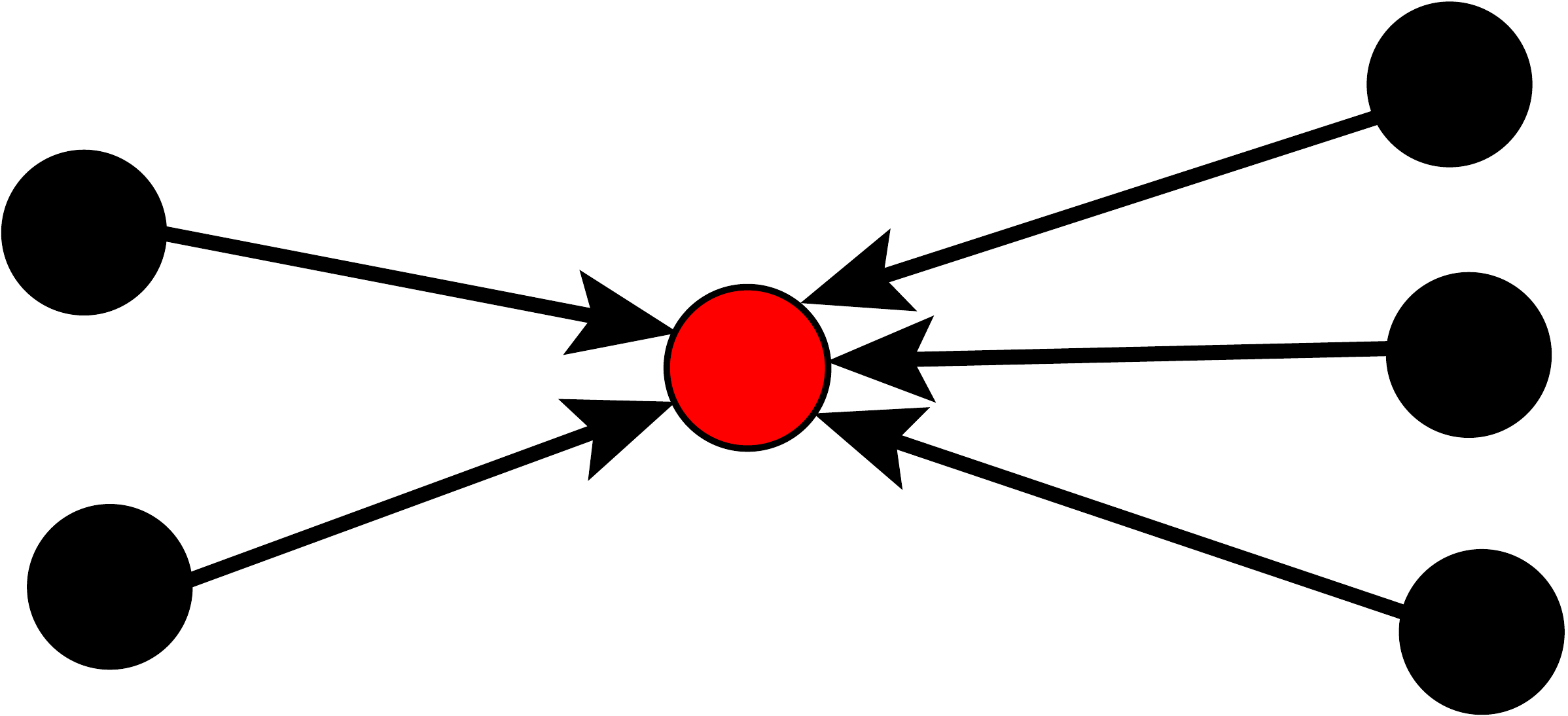} &
\includegraphics[width=2cm]{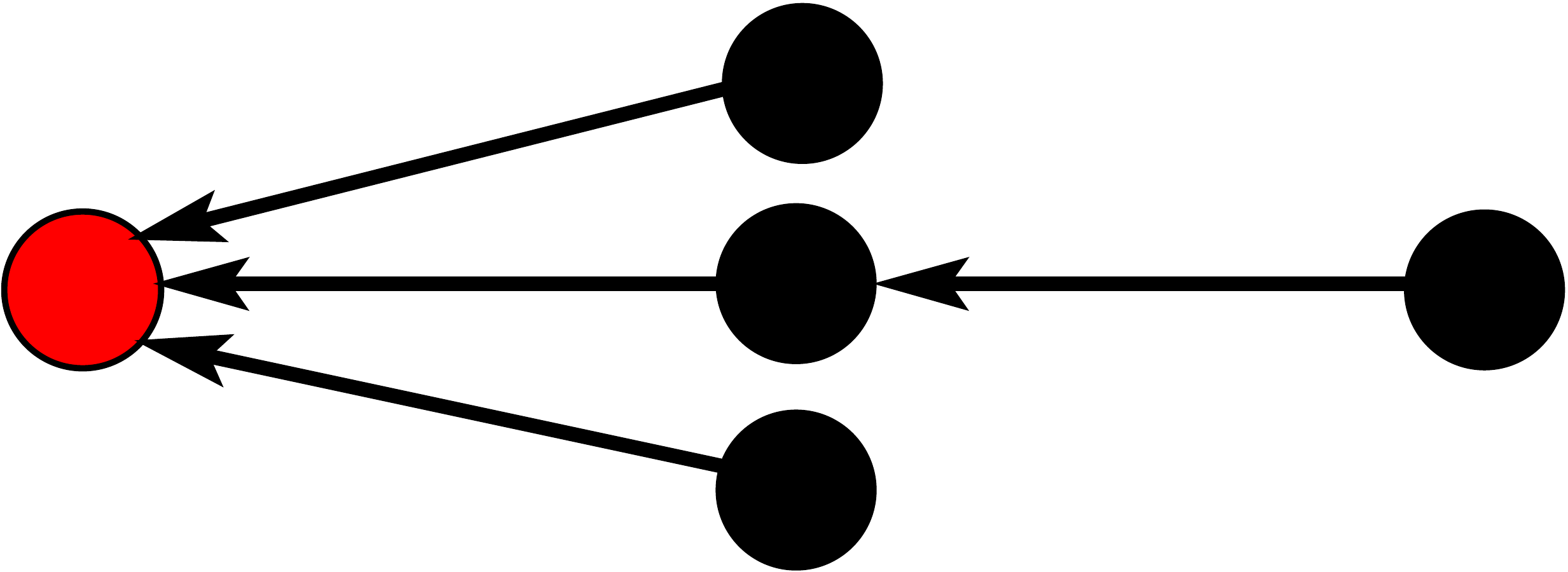} &
\includegraphics[width=2cm]{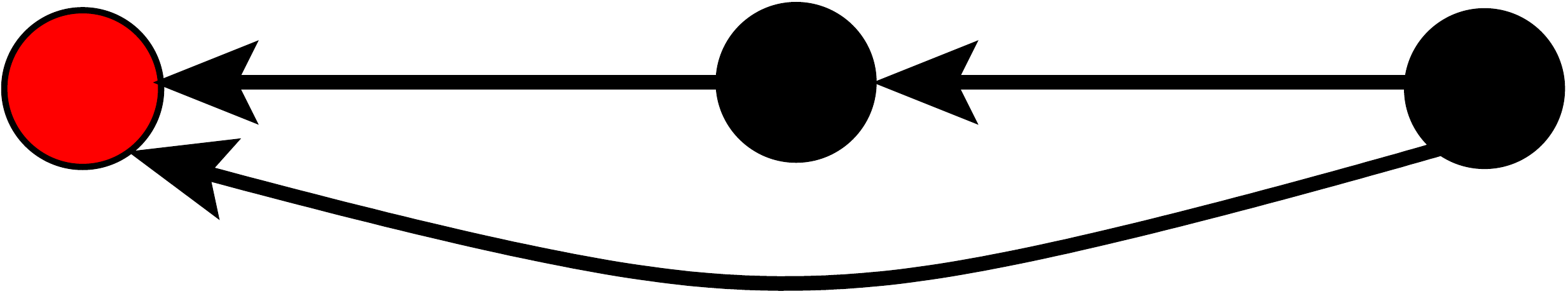} &
\includegraphics[width=2cm]{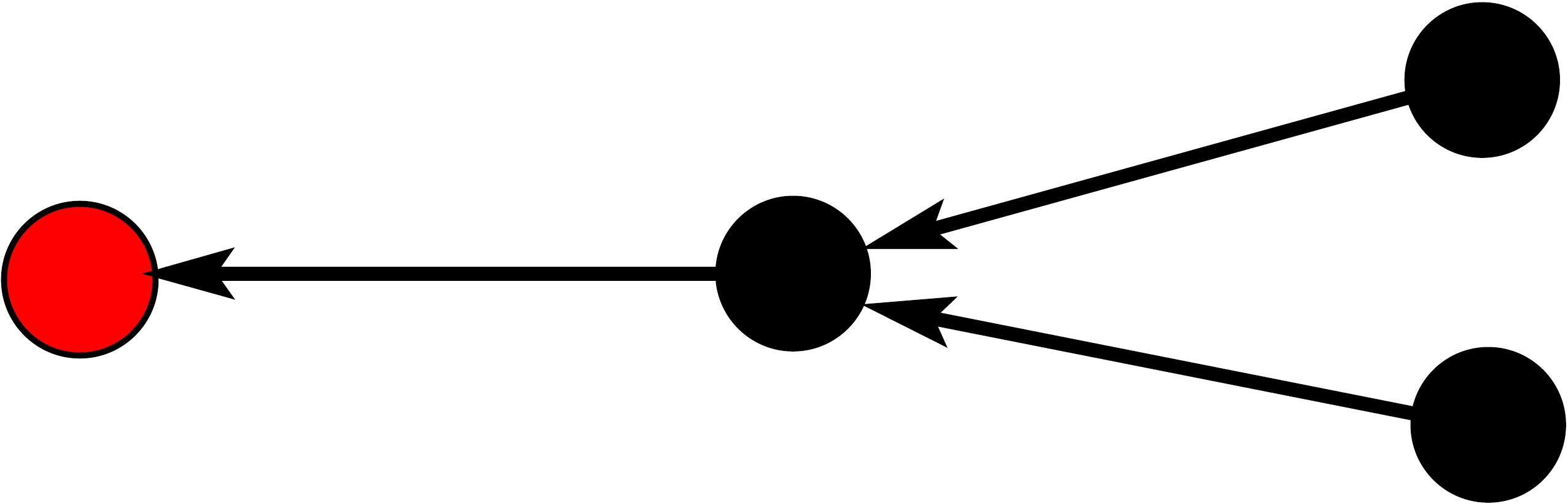} \\
\hline
Frequency & 134 & 83 & 56 & 52 & 46 \\
\hline
\end{tabular}

\vspace{2mm}

\begin{tabular}{|>{\centering}m{2cm}||>{\centering}m{2cm}|>{\centering}m{2cm}|>{\centering}m{2cm}|>{\centering}m{2cm}|>{\centering\arraybackslash}m{2cm}|}
\hline
Rank & 11 & 12 & 12 & 14 & 15  \\
\hline
Rank in \cite{leskovec2008thesis} & 9 & 13 & 14 & 15 & 12  \\
\hline
Shape & 
\includegraphics[width=2cm]{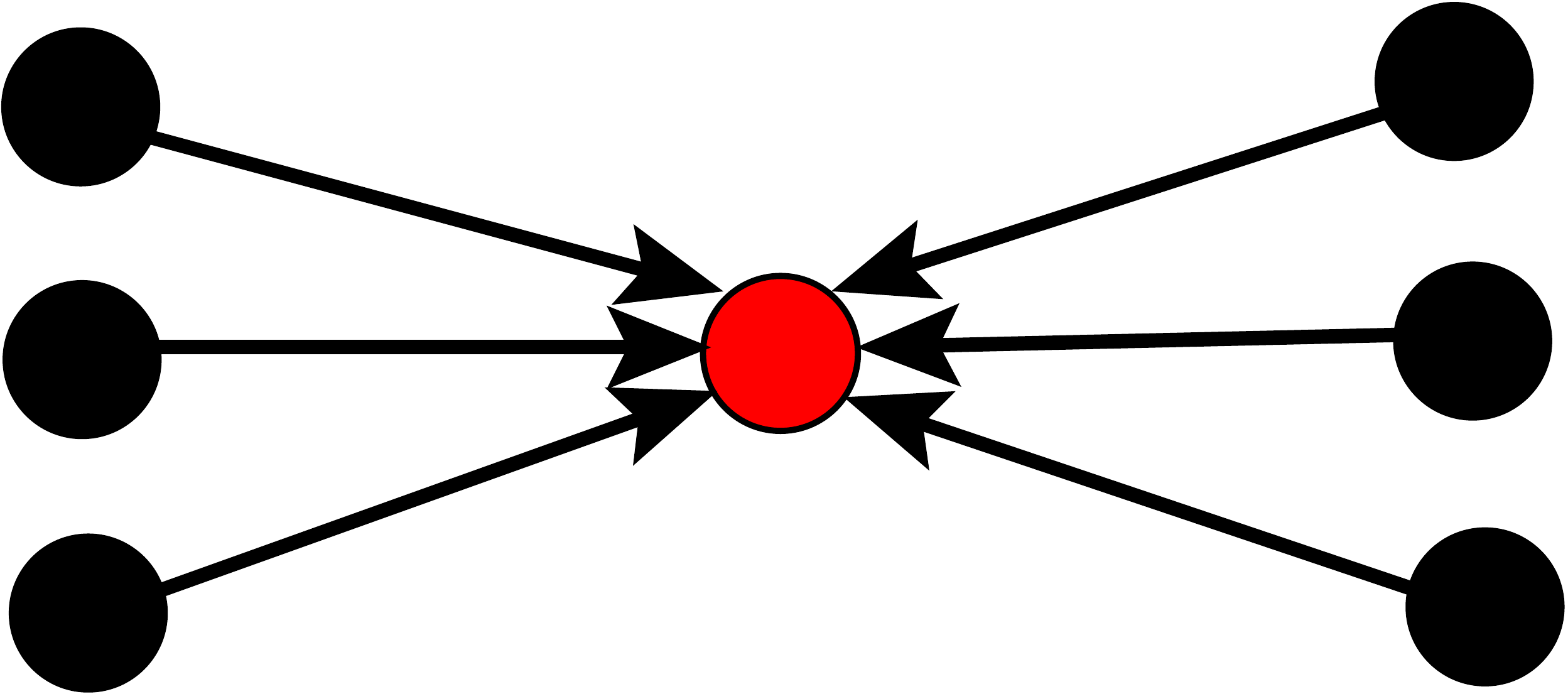} & 
\includegraphics[width=2cm]{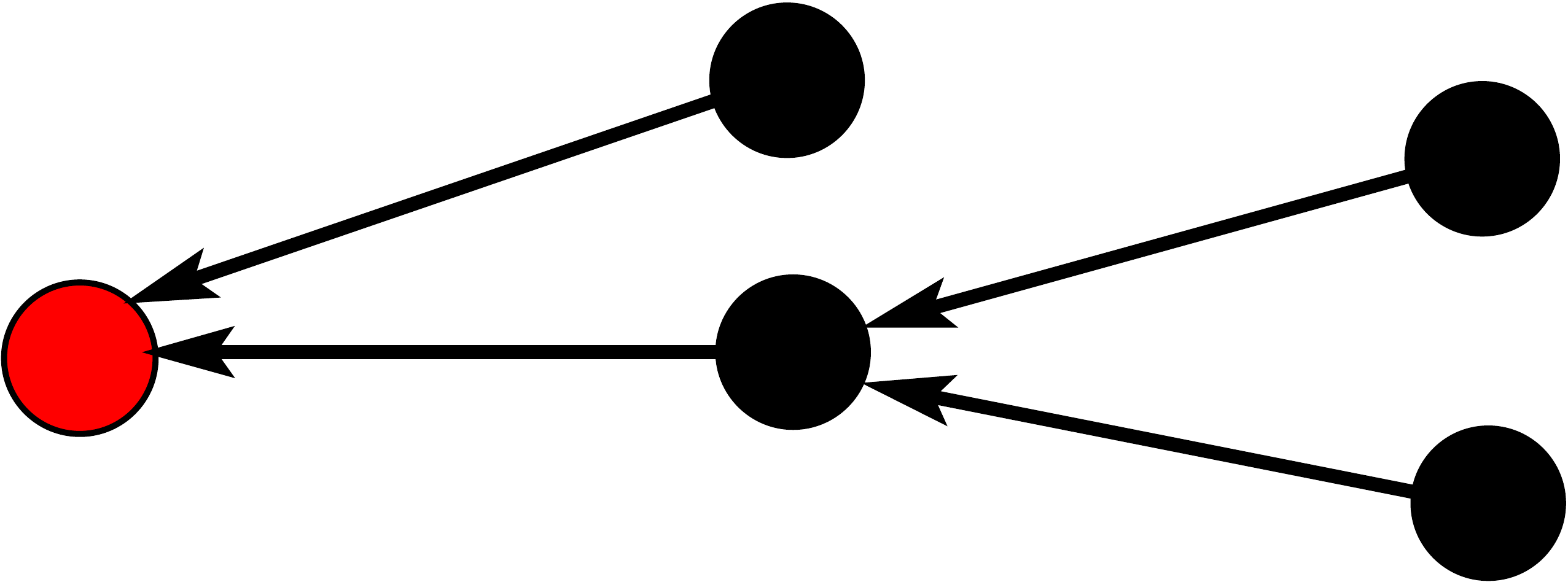} &
\includegraphics[width=2.1cm]{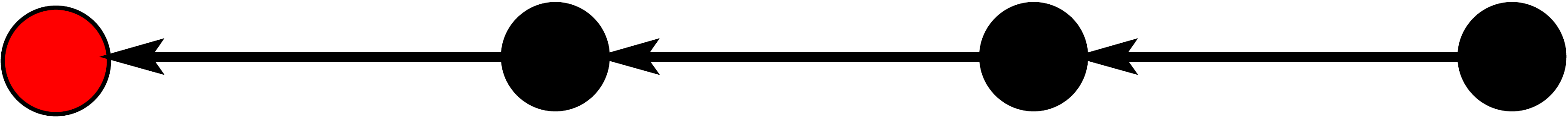} &
\includegraphics[width=2cm]{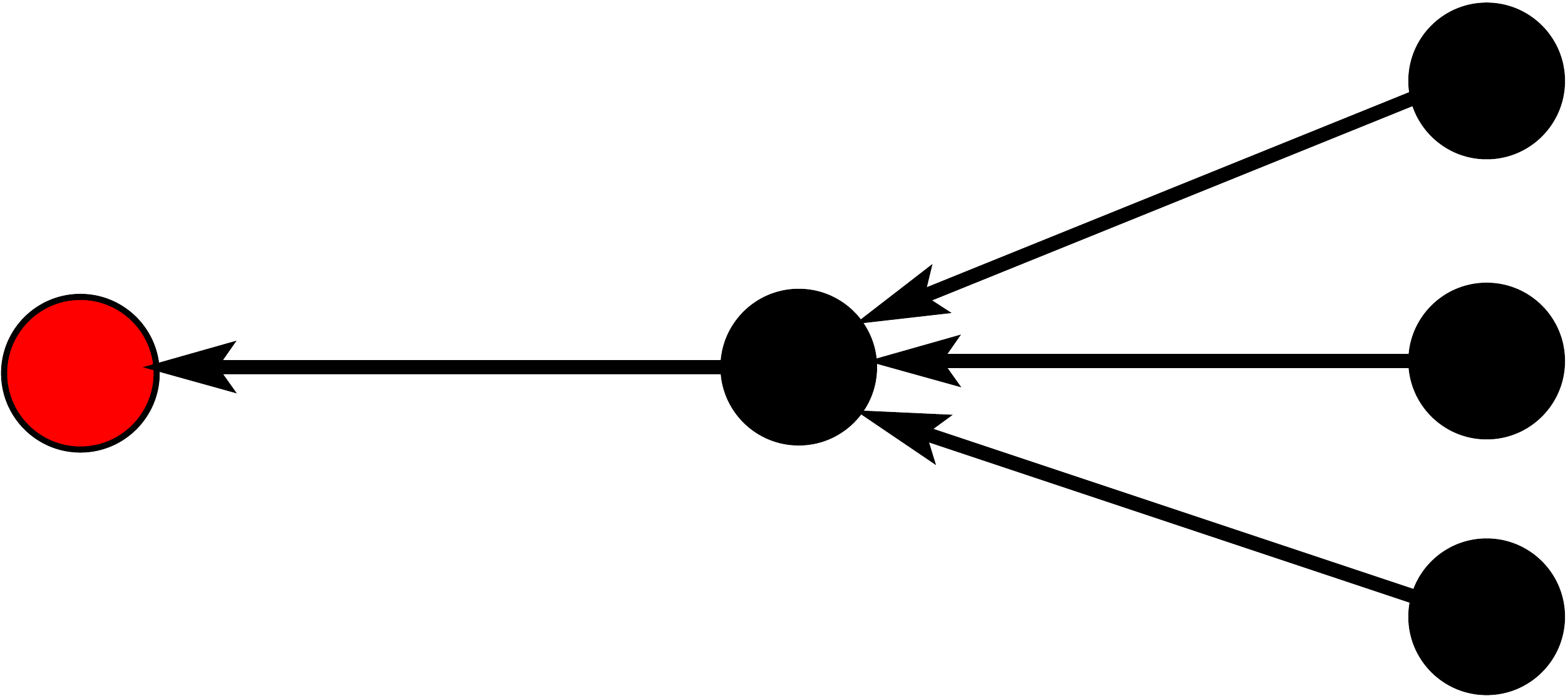} &
\includegraphics[width=2.1cm]{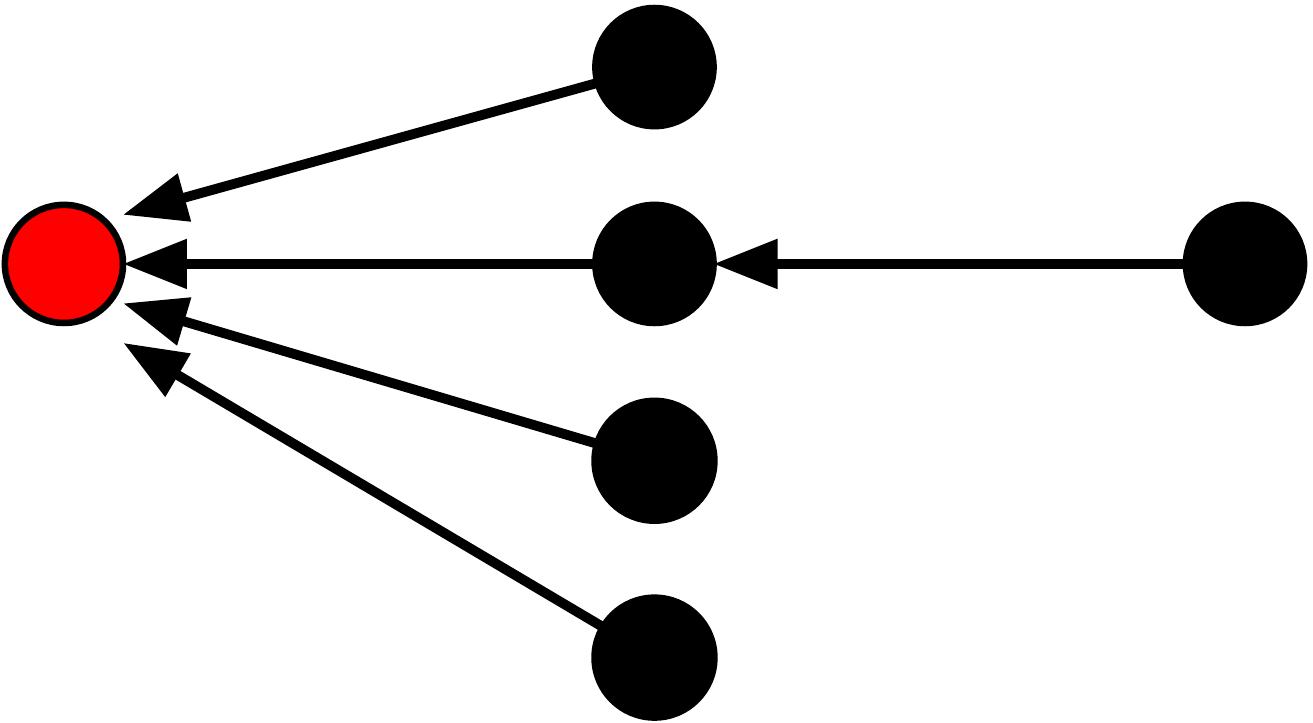}  \\
\hline
Frequency & 33 & 30 & 30 & 29 & 28\\
\hline
\end{tabular}

\end{center}
\caption{\label{tab:shape_casc} Cascade shapes ranked by frequency.}
\end{table*}

\subsubsection{Comparative summary}

In Table~\ref{tab:comparative_res}, we summarize the results reported above and compare them to the values in \cite{leskovec2007cascading}.
It enlightens striking similarities between the datasets despite the differences in origins and sizes.
These observations bring us to think that the underlying process of cascade building may be similar.
However, this is not necessarily evidence of the social mechanism of information spreading in this media: this may be the mere product of statistical properties of the network itself, an issue that we intend to address in the following.

\begin{table*}[ht]
\begin{center}
\begin{tabular}{|c||c|c|c|c||c|c|c|c|c|}
\hline
Dataset & Duration & $B$ & $N$ & $L$ & $r$ & $\alpha$ & $\beta$ & $\tau$ & $\gamma$ \\
\hline
\textit{Webfluence} & 151 days & 3,199 & 461,134 & 20,885 & 0.18 & -1.60 & -1.60 & -1.50 & -2.10 \\
\hline 
in \cite{leskovec2007cascading} & 90 days & 44,362 & 2,422,704 & 245,404 & 0.16 & -1.70 & - & -1.60 & -1.97 \\
\hline
\end{tabular}

\end{center}
\caption{\label{tab:comparative_res}
Comparative results between various datasets.
$B$: number of blogs,
$N$: number of posts,
$L$: number of citations,
$r$: Pearson correlation coefficient between number of in- and out-links of nodes.
Fitting with power-law models, we report the following exponents:
$\alpha$: blog in-links distribution, 
$\beta$: blog out-links distribution,
$\tau$: latencies distribution,
$\gamma$ : cascade sizes distribution.
}
\end{table*}

\section{First insights from a content-based analysis of large cascades\label{sec:first_insights}}

For an \emph{item} to exist, it would be expected that the posts of a cascade deal with the same topic; however, identifying the topic of a cascade is a hard task.
A standard method calls to the use of text mining and natural language processing techniques as well as strict statistical criteria such as the ones developed in \cite{gruhl2004information,adar2005tracking}; however the experimenter will still be required to make assumptions on parameters.
Besides, it has been underlined that blog contents are difficult to process with such techniques, e.g., \cite{joshi2007web}.
Even if our dataset is cleaned out from spam blogs and advertisements, we have to cope with semantic data with few structural conventions, as well as complex content, like lingo or hidden meaning.

Here, we want to get an overview of what different posts in a same cascade deal with.
A manual investigation provides us with the possibility to do so, in a more controllable way than what automatic tools allow.
%
%
We randomly chose 50 cascades among the top 5\% larger cascades (in terms of number of posts), that is to say 50 of the 526 cascades involving 9 nodes or more.
We draw benefit from this relatively low amount to achieve a qualitative inspection and get some insights on the real processes involved in citation cascades.

\subsection{Topic of a cascade \label{subsec:topic}}


Our goal here is to perform a manual inspection of large cascades to define topics in a flexible way, which will give us insights to take our analysis one step further.
The basic philosophy is to define a topic large enough to be shared by as many posts as possible, but restricted enough to be specific to the cascade.
The so-called topic is identified according to the following:
\begin{itemize}
\item
A topic is defined as an expression or set of expressions which can be considered as addressed in a post, it is not necessary that the expression itself appears in the text (this is one advantage of the manual investigation).
%

\item
From a practical point of view, after reading the posts, we establish a list of expressions that may be the most widely spread in the cascade, and which are not part of the set of existing topics.
Then for each post, we decide if it deals or not with each expression; the topic is the one which is shared by the largest amount of posts.

%
%
%
%

\end{itemize}

For example \textit{Nicolas Sarkozy} is a very usual topic of the French blogosphere during this period.
Now, if we demand that the posts refer to \textit{Nicolas Sarkozy \& regional elections}\footnote{In France, a \textit{r\'egion} is the largest administrative subnational division, the council ruling them is elected every 6 years and the latest elections took place in March 2010, i.e., during the crawling period.}, we narrow the topic down, but still several cascades may be defined by this expression.
But there is only one large cascade defined by \textit{Nicolas Sarkozy's comments on the regional elections results}.
Other examples of topics include: \textit{Death penalty in the US: Hans Skinner case}, \textit{Novelties revealed during Facebook conference} or \textit{The Geneva international motor show}.

Obviously, some choices remain arbitrary, for example it may be tricky to settle whether a post actually deals with a topic, and other choices lead to slightly different estimations.
However, we tested a strict versus a flexible convention on the sample, the former leading to lower topic-unity (see Sec.~\ref{subsec:tu.vs.sc}) evaluations than the latter, which is implemented in the following; this choice does not significantly affect our conclusions.
%

\subsection{Topic mutations along cascades}

The manual analysis presented in Section~\ref{subsec:topic} strengthens our daily experience intuition that the topic of a citing post may be quite different from the cited post's.
We develop this idea more quantitatively in the following by means of measures connecting structural features to the post content.

\subsubsection{Star-shaped and chain-shaped cascades}
\begin{figure}[h!]
\begin{center}
\includegraphics[angle=0,width=.85\linewidth]{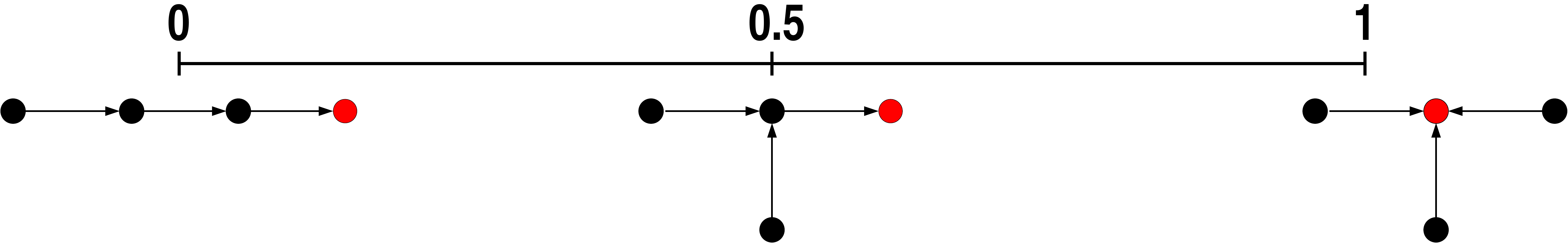}
\end{center}
\caption{\label{fig:star2chain} Behavior of $sc$ coefficient for examples of 4-node cascades.}
\end{figure}

In this section we propose a metric, denoted $sc$, to seize to what extent a given DAG belongs to the star vs chain topology.
Let $ n_o(x) $ and $ n_i(x) $ be the number of arcs going respectively from and to the node $x$ of the cascade $ \mathcal{C}$, then:
$$ sc = \frac{\sum _{x \in \mathcal{C} : n_i(x) = 0} n_o(x) -1 }{\sum _{x \in \mathcal{C}} n_o(x) -1 } $$
%
%
Long chains imply that nodes with outlinks also have inlinks, so that $sc$ is smaller when compared to cascades with the same amount of arcs, but with shorter chains.
This measure has been chosen to be normalized, so that chain-like patterns have $sc = 0$ and star-like $sc=1$, as summarized on Figure~\ref{fig:star2chain}.
%
%
Several other definitions may be suggested in these lines, however this one gives a good grasp of the feature we want to account for in the context of cascades.

\subsubsection{Topic-unity vs. \textit{sc} coefficient\label{subsec:tu.vs.sc}}
Topic-unity of a cascade is defined as the ratio of nodes in the DAG dealing with the topic attributed in accordance with the protocol of~\ref{subsec:topic}\footnote{Some posts were not online anymore when we checked their content; they are not taken into account for the topic-unity evaluation.}.
Here again, other definitions may be suggested (e.g. taking into account the fact that there may be more than one dominating topic in a cascade), however this estimate gives a first insight on the content of a cascade. 

We ranked the sample of cascades by increasing order, using the $sc$ coefficient on the one hand ($R_1$), and the topic-unity on the other hand ($ R_2 $)\footnote{We use rankings of $sc$ and topic-unity values as they are not homogeneously distributed on $[0;1]$.}.
Both rankings are correlated: the Pearson correlation coefficient $ r(R_1,R_2) = 0.57$, meaning that star-shaped cascades are more likely to exhibit a largely shared topic than chain-shaped ones.
In other words, the topic of a cascade is likely to change along the citations.
As the existence of an information item supposes that the posts deal with a unique topic, this analysis challenges the relevance of the notion of item in this context.
Notice that this observation is sociologically consistent: as the sustainability of a blog partly stems from the originality of its content, bloggers have a strong incentive not to copy-paste information that they found elsewhere, but rather to add their ``personal touch''.

\subsubsection{A glimpse of typical user behaviors}
To give the reader a more comprehensive grasp on the cascade content, we gather here some qualitative remarks with practical examples.

\begin{itemize}
\item 
Our manual investigation reveals that the blogosphere is flooded with information from external sources; in particular, references to mass media are extremely usual\footnote{In \cite{myers2012information}, the authors attempted to take this phenomenon into consideration in the context of Twitter.} --- especially, but not only, in the political blogosphere.

\item 
Cascades with high topic-unity (close to 1) often deal with the blogosphere itself, for example the set-up of a meeting of the far-left blogger community, or the reactions to a libel action against a blogger. 
Among most noticeable examples, organizing events such as games among bloggers is a usual practice, generating star-like shaped cascades (see Fig.~\ref{fig:game_casc}).
In this case, a picture or a motto is often used to refer to the event --- which can be seen as an item spreading.

\begin{figure}[h!]
\begin{center}
\includegraphics[width=.45\linewidth]{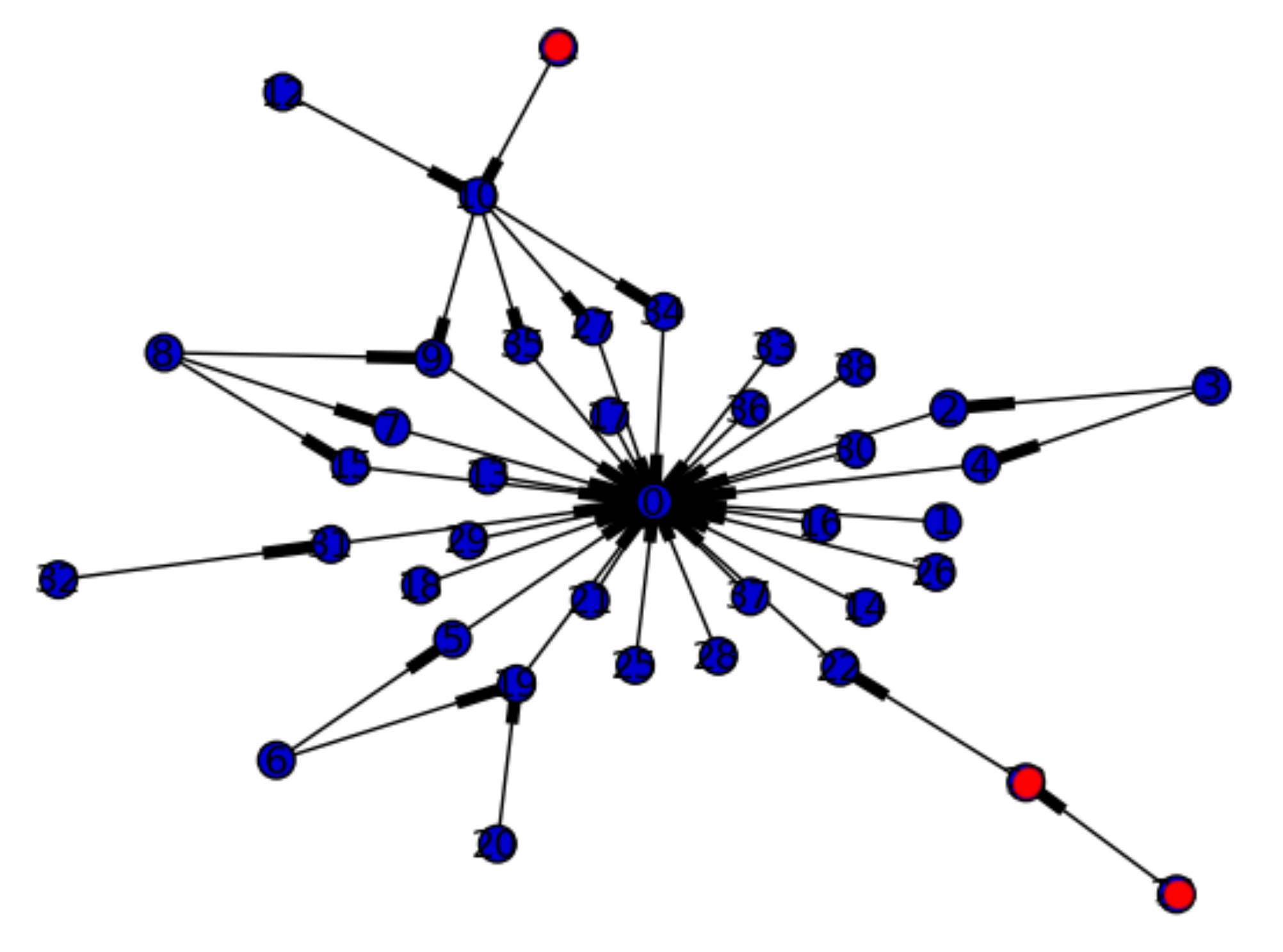}
\hspace{2mm}
\includegraphics[width=.45\linewidth]{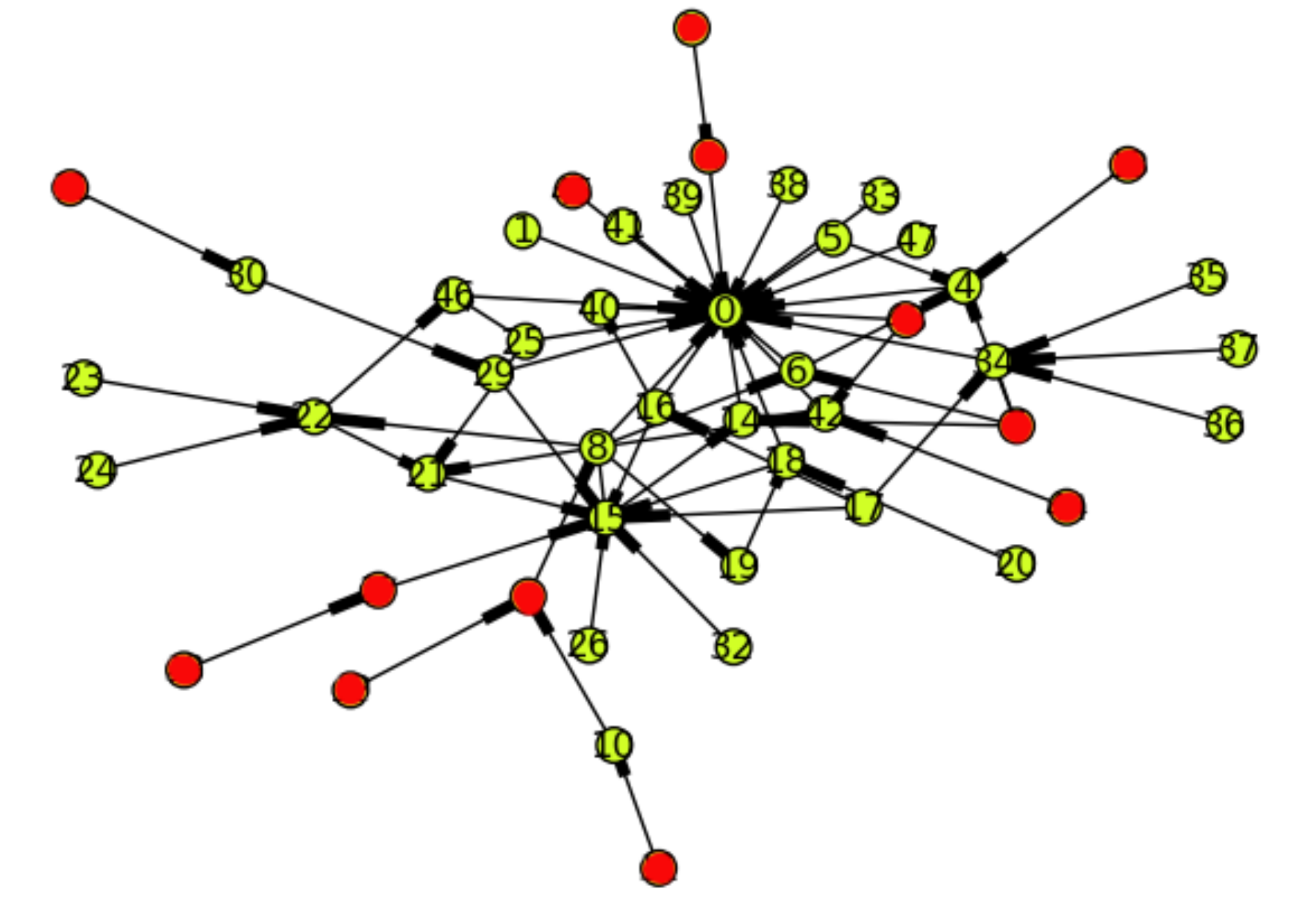}
\end{center}
\caption{\label{fig:game_casc}
Examples of high topic-unity cascades: red nodes do not deal with the topic of the cascade, or their content is no longer available.
Left: \textit{Cooking game: an olive-oil based sweet recipe}.
Right: \textit{About the law challenging anonymity on the Internet}.
}
\end{figure}

\item 
It can be observed that topic changes along a cascade are mostly brutal.
For example, the cascade \textit{Outcome of the `No-Sarkozy Day'} contains a secondary topic: \textit{3~Suisses sexist advertisement}, as can be seen on Figure~\ref{fig:mutating_casc} and the joint node reveals where and how the topic mutates. 
In this case, the citing blogger refers to the cited one as a representative of female-blogger community but without any relation to the content of her cited post, thus feeding the idea that some citations do not aim at spreading content but at acknowledging another blogger's status.
%
%
\begin{figure}[h!]
\begin{center}
\includegraphics[width=.6\linewidth]{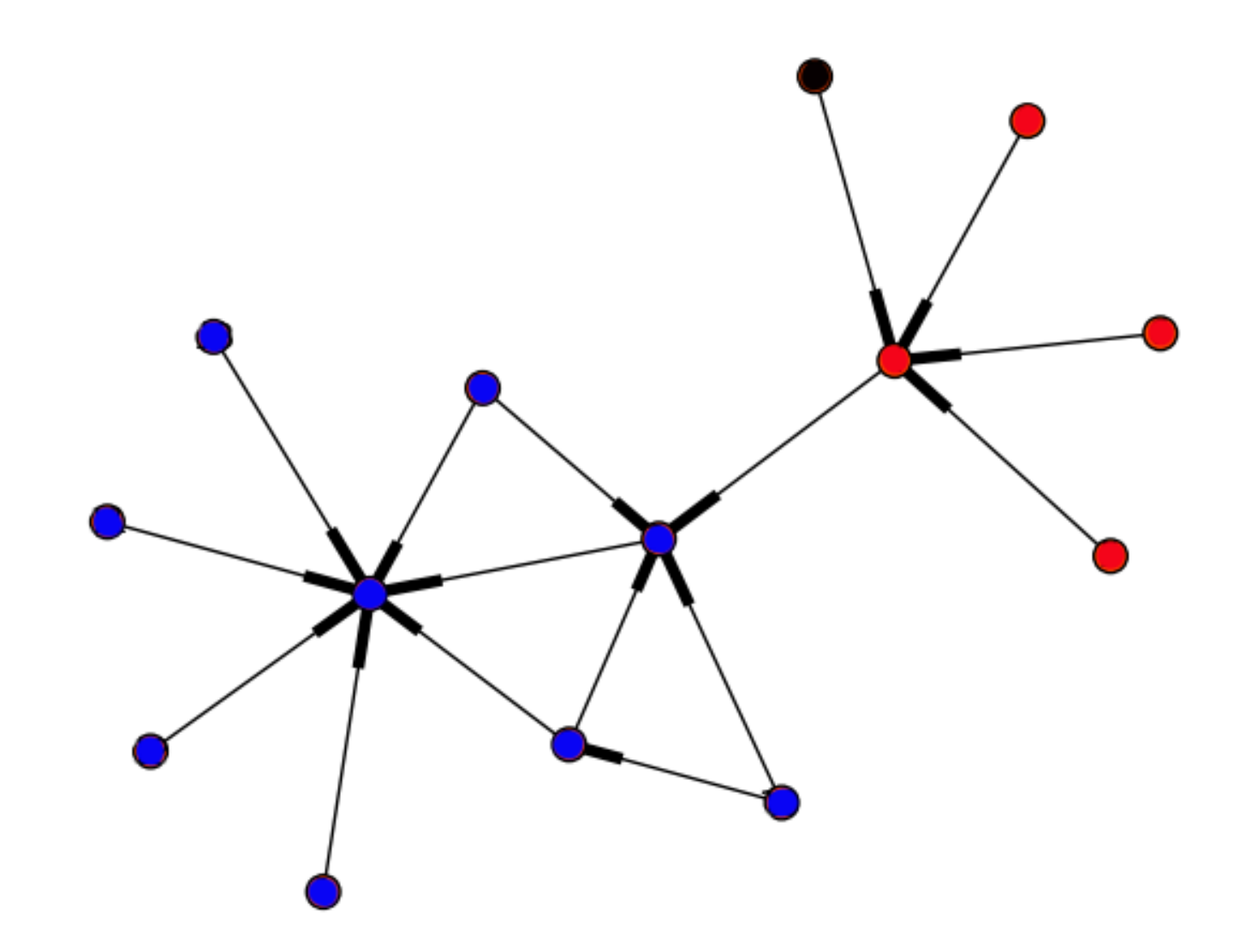}
\end{center}
\caption{\label{fig:mutating_casc}
Example of mutating topic. Blue nodes topic: \textit{Outcome of the `No-Sarkozy Day'}; red nodes topic: \textit{3 Suisses sexist advertisement}; black node: other.
}
\end{figure}

\item 
The topic-unity of cascades being questioned, it is not surprising to observe that the origin of a cascade, which is unique according to our definition, only rarely defines its topic.
The ability of a node to trigger a discussion, e.g., \cite{papagelis2009information} should thus be adapted to a formalism where a cascade size might be an inadequate measure of the intensity of a discussion.

\item 
Depending on the domain of interest, usual writing behaviors are clearly different.
For example, in the political blogosphere, the average post is quite long and often deals with several topics, it implies a higher topic mutation probability than posts dealing with individual hobbies, as cooking or knitting.

\item
In the same line, bloggers may have very different behaviors regarding their use of citations.
Some give an overview of their community in the form of an heterogeneous review, other rather cite with regard to the very specific topic they are dealing with\footnote{For examples illustrating these various behaviors, see respectively:
\url{www.geeek.org} and \url{falconhill.blogspot.fr}.}.
In~\cite{mcglohon2007finding}, the authors measured that typical cascade type strongly depends on the community it belongs to; this observation and ours both suggest that the explanation is rooted in the typical citing habits of bloggers in a community.


\item
Citations are frequently associated to comments on other bloggers, which give evidence of tight relationships : regular readers, sometimes off-line friends etc.
We could therefore think of detecting communities using cascade analysis.
For example, it seems likely that a group of bloggers participating to several cascades of unrelated topics may be part of the same social group on the web.

\end{itemize}

\section{Modeling cascades without items \label{sec:model}}

Models describing citation data often call to an epidemic-like description of spreading \cite{adar2005tracking,gotz2009modeling}.
Hence, they aim at reconstructing some features of the observed citation structure using a detailed description of the behaviors of users in the network.
While it may happen that a piece of information is duplicated without any change from a blogger to another, a closer examination of the content show that this behavior is not dominant.
%
%
The analysis of Section \ref{sec:first_insights} makes us think that different citations within the same cascade may not be strongly correlated.
In this section, we propose a simple model consistent with this observation, that is to say independent from items, and show that it is sufficient to reconstruct most features of the cascades.


\subsection{Model description and results}

The central idea of this model is that citations may be considered without any reference to any item spreading on the network, so that citations are treated as much as possible as uncorrelated events.
We thus focus on the structure of the underlying network and on the citing activity of bloggers to account for the citation cascades observed.

We built a very simple model according to the following description: 
\begin{itemize}
\item If a post $P_a$ of blog $a$ refers to a post $P_b$ (of $b$), we now consider that $P_a$ cites \textit{any} post already published by $b$.
\item The post cited is selected randomly, but with a probability bias that enables to fit the real latency distribution. 
In more details, the process consists in choosing a post randomly, and possibly discarding it according to a power-law function of the latency; the parameters of the power-law are set to fit the distribution on Fig.~\ref{fig:cit_dyn}.
\end{itemize}
Without this second prescription, the distribution of latencies is much alike the one obtained for a poissonian process, exhibiting in average longer lapses of time between citations.
The number of posts cited being roughly 2\% of the total number of posts published in the dataset, this randomization process is supposed to break efficiently correlations between the events of a cascade.\\

%

%
%
%
%

\begin{figure}[h!]
\begin{center}
\includegraphics[angle=-90,width=.49\linewidth]{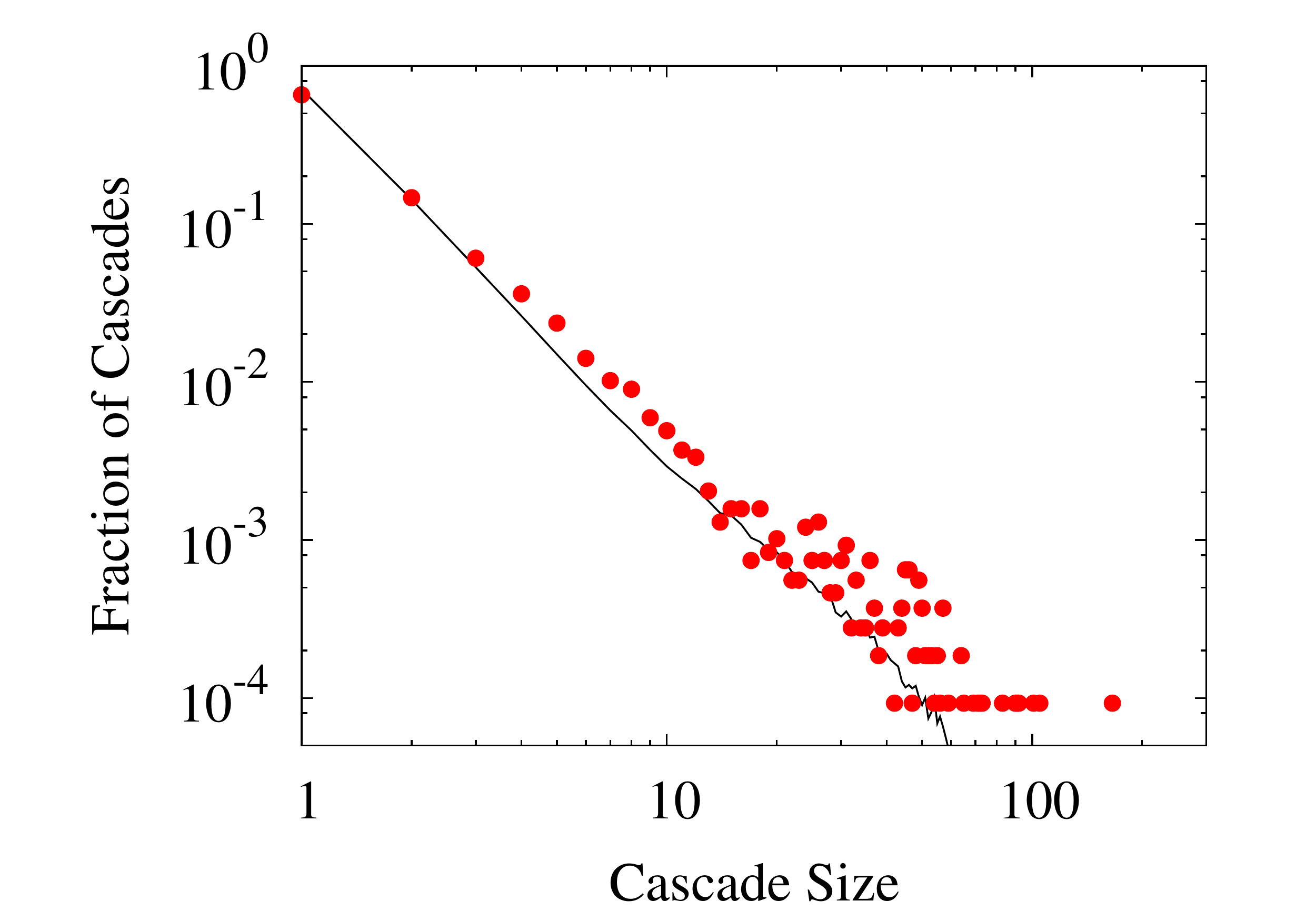}
\includegraphics[angle=-90,width=.49\linewidth]{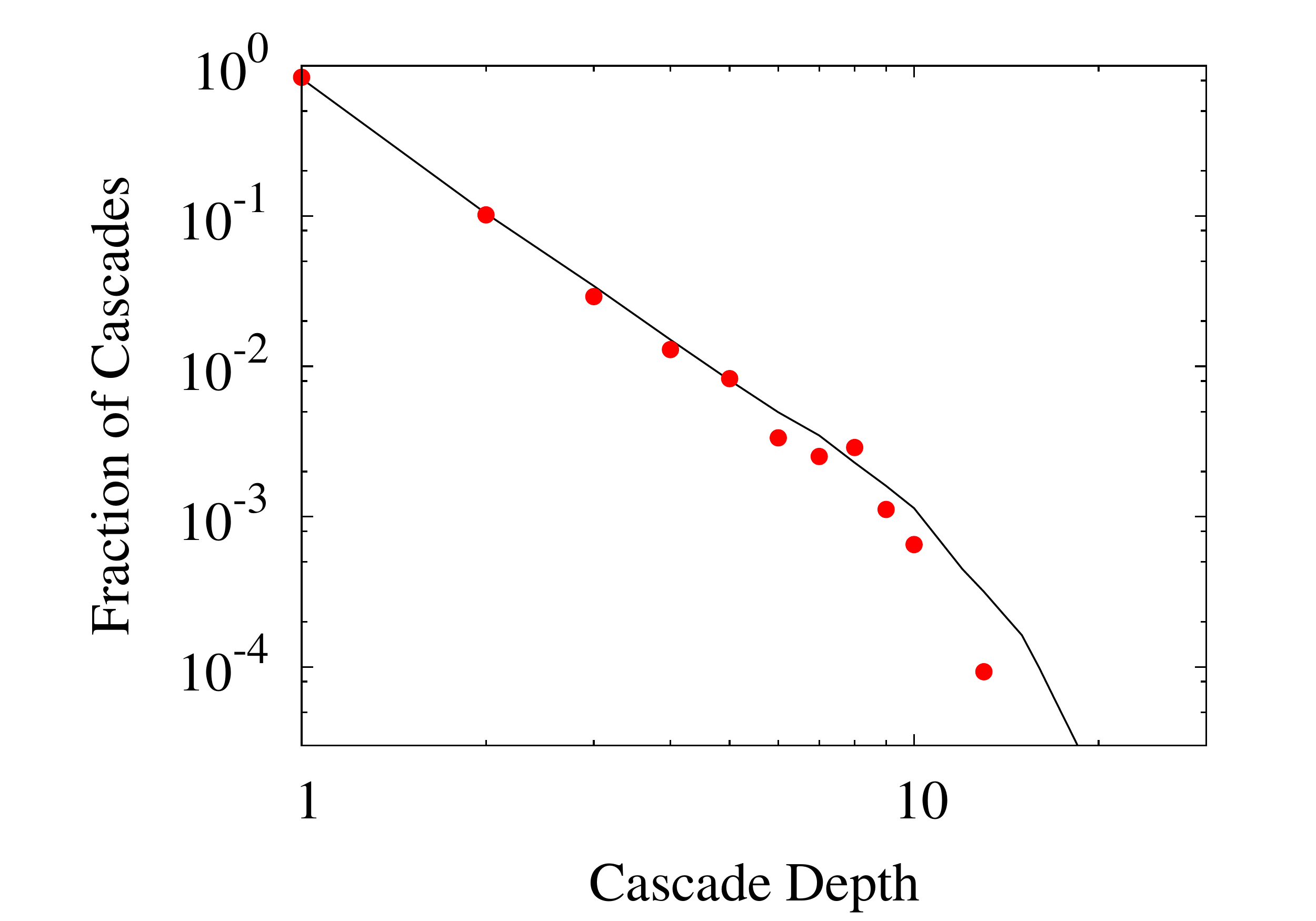}
\end{center}
\caption{\label{fig:casc_size_comp}
Comparison between real data (red circles) and an average of 100 realizations of the model (black line). 
Left: probability distribution of cascade sizes.
Right: probability distribution of cascade depths.
}
\end{figure}

We measure statistical features of the data generated and compare them to the original dataset (see Figure~\ref{fig:casc_size_comp}).
First, we observe that sizes and depths follow heterogeneous distributions, and more precisely a power-law model seems appropriate to fit the size distribution.
Both the size and depth distributions of the model are close to the real one, except for slightly smaller cascades (fits give a 2.3 slope for the model, versus 2.1 for the real data).
So in spite of its great simplicity, this model is able to mimic these important features of the original dataset.

We then compare in more details the patterns found in our model to real data.
\begin{figure}[h!]
\begin{center}
\includegraphics[width=1.05\linewidth]{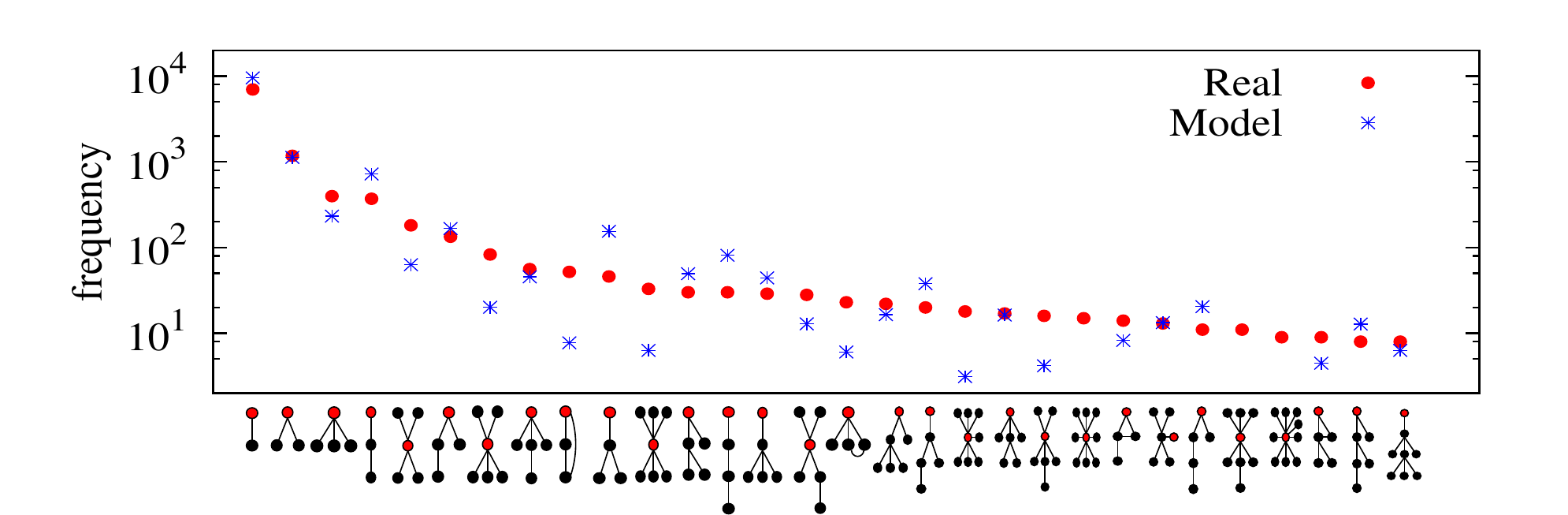}
\end{center}

\caption{\label{fig:comp_motif} Frequency of most usual cascades. The model is averaged over 100 realizations, values absent from the picture are below the minimum of the frequency range (=2).}
\end{figure}
Figure~\ref{fig:comp_motif} reveals that the number of patterns produced by the model is roughly of the same order of magnitude as in the original dataset.
However, rankings are not identical, and there are significant differences between the real data and our model.
Two types of cascades are clearly underestimated by the model (in some cases even nearly absent):
the first exhibits several outgoing links from a node to other nodes of the cascade, and consequently displays cyclic patterns:~\includegraphics[width=2cm]{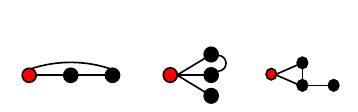}.
The second type represents star-like cascades:~\includegraphics[width=3cm]{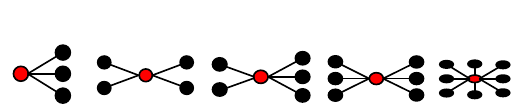}.
More generally, it can be observed on Figure~\ref{fig:comp_motif} that most underestimated cascades have high $sc$ values (typically $sc \in \left[ \frac{2}{3};1 \right] $) while overestimated cascades have lower $sc$.
As discussed in Section~\ref{sec:first_insights}, star-shaped cascades may exhibit high topic-unity, while the random cascade generator is blind to such semantic effects.
We may then assume that a fraction of cyclic and star-like cascades result from correlated citations, and therefore cannot be rebuilt by the model.
In other words, these patterns are salient characteristics of the real dataset.
It can be related to other works ~\cite{zhao2010communication,tabourier2012detect}, in which the authors look for evidence of spreading phenomena on other types of large communication networks by comparing the number of dynamical patterns in real data to various dynamical benchmarks.


\subsection{Discussion}

The model described above gives clues that some citation patterns are more likely than others to carry information spreading --- namely, cyclic and star-shaped cascades.
Even in this case, it is more accurate to talk about correlated citations rather than information spreading; this assumption is supported by the former semantic analysis.
As the model merely takes into account who cites who and when, we think that the structure of the cascades observed is mostly due to the structure of the ego network of bloggers when it comes to citation.
From this perspective, citations often seem loosely related to the intent of spreading an information.
A possible explanation is that a substantial fraction of citations may rather be a mean to acknowledge an acquainted blogger.

The model presented above is very simplistic, and we do not claim that it fully captures the mechanisms of citing behaviors on the blogosphere.
Yet, the fact that such a model rebuilds some important features of cascades implies that we should be very cautious when using any other model in this context.
More precisely, distributions of citation cascades sizes and depths are clearly not reliable traces of diffusion phenomena; the number of cascades per type may be more relevant in this context, but richer metrics are needed to prove information spreading.
As we can hardly infer more information from a purely structural analysis, it calls for the use of other metrics, like content or detailed temporal information.
These observations deter from using elaborate models describing complex agent behaviors: we would not be able to know if the outcome stems from the assumptions, or only from the lack of relevant measure to settle if the synthetic data resembles the real one.
It raises tricky questions: what is a good benchmark to compare real data with? And consequently, what is an \textit{expected property} in the context of blog citation cascades?


\section{Conclusion}

In this paper, we tackled the problem of the origin of citation cascades in blog networks.
Measurements indicate that our dataset shares many statistical properties with the one in~\cite{leskovec2007cascading}, while focusing on an unrelated subpart of the blogosphere.
We then put forward the assumption that the underlying process may be identical in both cases.
As epidemic-like models are popular in this area, we explored the contents to see if we can identify items, i.e., propagating piece of information, throughout a cascade. 
But items are difficult to define in this specific context: we made the minimal choice of isolating a common topic of discussion, and even in this case we observed that topics often change along a cascade.
A simple model based on independent citations, turns out to mimic well some features of the real dataset, namely the size and depth distributions.

It brought us to reconsider the results obtained with models in this context: if such a basic model rebuilds these observations, these features do not reveal information spreading. 
According to an epidemic-like description of information propagation, it can be compared to a virus spreading among a population while mutating.
However, here we have no simple equivalent of the genetic distance, and even if we did, we should probably consider very high rates of mutation.
Our belief is that diffusion in the sense of an epidemic spreading is not appropriate to model what happens on this network.
%
%
In fact, we questioned the very idea that citations primary function is information spreading.

Another objection to an epidemic-like description comes from the fact that it relies on the assumption that the network is the only medium of propagation.
But even if bloggers are known to be very active consumers of online content~\cite{lenhart2006bloggers}, there is no guarantee that the information spreading comes from the blogosphere itself.
Content and structure analyses reveal that the blogosphere is flooded with information from external sources: the flow generated by mass media, other social networks, and off-line experience are presumably more influential on the posting activity than other blogs.
It is consistent with the sociological literature, which describes information adoption as a complex interplay between mass media and local opinion dynamics \cite{katz1957two}. 
Furthermore, recent works acknowledge the existence of external sources of information in the context of online social media, observing ``jumps'' of information and trying to both measure and model this phenomenon \cite{myers2012information}.
In that sense, the epidemiological metaphor may be misleading when transposed to social media.

Contagion models are attractive as they propose a simple, versatile way to describe subtle processes; besides, it is a fact that word of mouth spreading of information on social media does sometimes happen, but that does not make it a measurable and/or dominating means of information spreading.
Such description has little large-scale empirical support, as tracking spreading in social media is a very challenging task.
%
%
Our study deals with specific patterns of a particular family of datasets, other fields may be more suited to epidemic tools.
We pointed out a few ingredients that would contribute to more reliable analyses: a strict definition of the item spreading in the system, and a network where sources of information are identified and controllable.
Efforts should also focus on defining an adequate set of measurements to locate information spreading traces in blog data, and more broadly in social datasets.
Cyclic and star-like patterns are unexpectedly numerous with regard to the benchmark that we proposed.
It suggests that they play a specific role in the \textit{Webfluence} dataset and could be good candidates to fingerprint correlated events in various contexts.

\section*{Acknowledgements}
We would like to thank Christian Borghesi and Renaud Lambiotte for their useful comments.

This work has been partially supported by the FNRS and the City of Paris \textit{Emergence} program through the \textit{DiRe} project.

The data has been collected as part of the French National Agency of Research \textit{Webfluence} project \#ANR-08-SYSC-009.

This paper presents research results of the Belgian Network DYSCO (Dynamical Systems, Control, and Optimization), funded by the Interuniversity Attraction Poles Programme, initiated by the Belgian State, Science Policy Office. The scientific responsibility rests with its authors.

\bibliographystyle{plain}
\bibliography{biblio_blog}

\end{document}